\documentclass[conference, letterpaper]{IEEEtran}
\usepackage{graphicx}
\usepackage{textcomp}
\usepackage[table]{xcolor}

\usepackage{graphics}
\usepackage[utf8]{inputenc}
\usepackage{cite}
\usepackage{color}
\usepackage{amsmath}
\usepackage{amssymb}
\usepackage{amsfonts}
\usepackage{mathtools}
\usepackage{bm}
\usepackage{pifont}
\usepackage{multirow}
\usepackage{hhline}
\usepackage{cellspace}
\usepackage{makecell}
\usepackage{listings}
\usepackage{soul} % Highlighting text
\usepackage{algorithmic}
\usepackage{pdfpages}
\usepackage{subcaption}
\usepackage{stmaryrd}
\usepackage{booktabs}
\usepackage{dashbox}
\usepackage{thmtools}

\def\BibTeX{{\rm B\kern-.05em{\sc i\kern-.025em b}\kern-.08em
    T\kern-.1667em\lower.7ex\hbox{E}\kern-.125emX}}

\newtheorem{lma}{Lemma}

\definecolor{darkgreen}{RGB}{0,127,0}
\definecolor{darkred}{RGB}{127,0,0}
\definecolor{shadecolor}{RGB}{250, 250, 250}
\definecolor{lightyellow}{RGB}{250, 250, 212}
\definecolor{HLYELLOW}{RGB}{240, 127, 0}
\definecolor{hlyellow}{RGB}{240, 127, 0}
\sethlcolor{white}

\usepackage{xcolor}
\colorlet{hlcolor}{yellow!20}

\DeclarePairedDelimiter\ceil{\lceil}{\rceil}
\DeclarePairedDelimiter\floor{\lfloor}{\rfloor}

\newcolumntype{P}[1]{>{\centering\arraybackslash}p{#1}}

\lstdefinestyle{cpp}{ %
	language=C,
	frame=single, 
	framerule=0pt,
	basicstyle=\small\ttfamily, 
	backgroundcolor=\color{shadecolor},
	keywordstyle=\color{blue}\bfseries,
	commentstyle=\color{darkgreen},
	rulecolor=\color{black},
	stringstyle=\color{darkred},
	lineskip=0pt,
	keywords={},
	numbers=none,
	numbersep=5pt,
	showstringspaces=false
}

\lstdefinestyle{tablestyle}{
    language=Python,
	frame=single, 
	framerule=0pt,
	basicstyle=\scriptsize\ttfamily, 
	keywordstyle=\color{blue}\bfseries,
	commentstyle=\color{darkgreen},
	rulecolor=\color{black},
	stringstyle=\color{darkred},
	lineskip=0pt,
	keywords={def, @dace, program, map, lambda, import, numpy, np, int32, float32, float64, complex128, with, as, True, False, return, for, in, dc, nb, numba, dace, cupy, cp, @numba, @nb, @dc, symbol},
	numbers=none,
	numbersep=5pt,
	showstringspaces=false
}

\usepackage[textwidth=0.7cm,colorinlistoftodos]{todonotes}
\usepackage{soul}
\soulregister\cite7
\soulregister\ref7
\soulregister\pageref7
\soulregister\lstinline7

\usepackage[normalem]{ulem} % use normalem to protect \emph

\definecolor{revisionhighlight}{rgb}{0.8, 0.8, 1}

\definecolor{sfcolor1}{HTML}{1b9e77}
\definecolor{sfcolor2}{HTML}{d95f02}
\definecolor{sfcolor3}{HTML}{e6ab02}
\definecolor{sfcolor4}{HTML}{e7298a}
\definecolor{sfcolor5}{HTML}{7570b3}
\colorlet{hightlightcolor1}{sfcolor1!30!white}
\colorlet{hightlightcolor2}{orange!20!white}
\colorlet{hightlightcolor3}{yellow!30!white}
\colorlet{hightlightcolor4}{sfcolor4!20!white}
\colorlet{hightlightcolor5}{cyan!20!white}

\DeclareRobustCommand{\revisionanotag}[1]{{\sethlcolor{white}\hl{#1}}}

\usepackage{tikz}
\usetikzlibrary{positioning,fit,calc,shapes.geometric}

\makeatletter % changes the catcode of @ to 11
\newcommand{\linebreakand}{%
  \end{@IEEEauthorhalign}
  \hfill\mbox{}\par
  \mbox{}\hfill\begin{@IEEEauthorhalign}
}
\makeatother % changes the catcode of @ back to 12

\title{\huge Deinsum: Practically I/O Optimal Multilinear Algebra}

\author{\IEEEauthorblockN{1\textsuperscript{st} Alexandros Nikolaos Ziogas}
\IEEEauthorblockA{\textit{Department of Computer Science} \\
\textit{ETH Zurich}\\
Zurich, Switzerland \\
alexandros.ziogas@inf.ethz.ch}
\and
\IEEEauthorblockN{2\textsuperscript{nd} Grzegorz Kwasniewski\IEEEauthorrefmark{1}}
\IEEEauthorblockA{\textit{NextSilicon} \\
Tel Aviv, Israel \\
grzegorz.kwasniewski@nextsilicon.com}
\and
\IEEEauthorblockN{3\textsuperscript{rd} Tal Ben-Nun}
\IEEEauthorblockA{\textit{Department of Computer Science} \\
\textit{ETH Zurich}\\
Zurich, Switzerland \\
tal.bennun@inf.ethz.ch}
\and
\linebreakand
\IEEEauthorblockN{4\textsuperscript{th} Timo Schneider}
\IEEEauthorblockA{\textit{Department of Computer Science} \\
\textit{ETH Zurich}\\
Zurich, Switzerland \\
timo.schneider@inf.ethz.ch}
\and
\IEEEauthorblockN{5\textsuperscript{th} Torsten Hoefler}
\IEEEauthorblockA{\textit{Department of Computer Science} \\
\textit{ETH Zurich}\\
Zurich, Switzerland \\
torsten.hoefler@inf.ethz.ch}
}

\usepackage[symbol]{footmisc}

\begin{document}

\maketitle

\begingroup
\renewcommand*{\thefootnote}{\fnsymbol{footnote}}
\footnotetext[1]{The author's affiliation at the time of submission was NextSilicon. However, a significant part of the research was done while he was affiliated with ETH Zurich.}
\endgroup

\begin{abstract}
Multilinear algebra kernel performance on modern massively-parallel systems is determined mainly by data movement.
However, deriving data movement-optimal distributed schedules for \revisionanotag{programs}
%applications
with many high-dimensional inputs is a notoriously hard problem.
State-of-the-art libraries rely on heuristics and often fall back to suboptimal tensor folding and BLAS calls.
We present Deinsum, an automated framework for distributed multilinear algebra computations expressed in Einstein notation, based on rigorous mathematical tools to address this problem.
Our framework automatically derives data movement-optimal tiling and generates corresponding distributed schedules, further optimizing the performance of local computations by increasing their arithmetic intensity.
To show the benefits of our approach, we test it on two important tensor kernel classes: Matricized Tensor Times Khatri-Rao Products and Tensor Times Matrix chains.
We show performance results and scaling on the Piz Daint supercomputer, with up to 19x speedup over state-of-the-art solutions on 512 nodes.

% Linear algebra kernels are the fundamental building blocks of virtually all scientific applications.
% Communication-efficient parallel implementations of those kernels are crucial in executing these applications at scale.
% These implementations depend, in turn, on methods that derive I/O lower bounds.
% Such a method is X-Partitioning and has been used to successfully determine tight I/O bounds for several linear algebra kernels, including matrix multiplication, Cholesky, and LU factorizations.
% This work presents a framework utilizing X-Partitioning to find asymptotically communication-optimal .5D decompositions for multilinear algebra kernels that operate on higher-order tensors.
% Based on these decompositions, we implement within our framework distributed algorithms for the matricized tensor times Khatri-Rao product (MTTKRP) and tensor times matrix chains (TTMc).
% We show performance results and scaling on the Piz Daint supercomputer, with up to X.XXx speedup over state-of-the-art solutions and up to XX.XX\% scaling efficiency on XXXX nodes.
\end{abstract}

% \begin{IEEEkeywords}
% % component, formatting, style, styling, insert
% \end{IEEEkeywords}

\section{Introduction}

Linear algebra kernels are the fundamental building blocks of virtually all scientific applications; from physics~\cite{plasma}, computational chemistry~\cite{cp2k}, and medicine~\cite{fmri}; to material science~\cite{omen}, machine learning~\cite{pytorch, tensorflow}, and climate modeling~\cite{gt4py,cosmo1,cosmo2}.
It is nigh impossible for any survey of the relevant scientific codes to not stumble at every step across multitudes of vector operations, matrix products, and decompositions from the arsenal of the ubiquitous BLAS~\cite{blas} and LAPACK~\cite{lapack} libraries.
Furthermore, the execution of these kernels often dominates the overall runtime of entire applications; and with current hardware trends, their performance is frequently limited by the 
data movement~\cite{berkeley} rather than FLOPs.
Therefore, the design of communication-efficient parallel algorithms for (multi)linear algebra is indispensable in efficiently executing the scientific applications at scale.

Linear algebra kernels, which operate on vectors and matrices, have been studied extensively.
There are a plethora of works on lower bounds and communication-avoiding schedules, e.g., for matrix multiplication~\cite{cosma, solomonik, optimalStrassen}, and matrix factorizations such as LU and Cholesky~\cite{conflux, choleskyQRnew}.
However, multilinear algebra, the extension of these methods on multidimensional arrays (higher-order tensors), is 
far less studied, especially in communication optimality.
The performance of critical computational kernels in data analysis, such as the CANDECOMP/PARAFAC (CP)~\cite{cppapers} and Tucker decompositions~\cite{tuckerpaper}, is largely untapped due to the complexity imposed by the high dimensionality of the iteration space. Although some works study the theoretical communication complexity of chosen multilinear kernels~\cite{ballard2018communication},
practical implementations tend to focus only on the shared-memory parallelization~\cite{gett, tensorGPU} due to the intrinsic complexity of efficient communication patterns.
To the best of our knowledge, the only broadly used library for distributed general tensor algebra expressible in the Einstein summation (einsum)
notation is the Cyclops Tensor Framework~\cite{ctf}.

% In general, linear algebra kernels operate on tensors of varying dimensionality.
% Low-order kernels utilizing vectors and matrices have been studied extensively, yielding lower bounds and communication-avoiding schedules for matrix multiplication~\cite{cosma, solomonik, optimalStrassen} and matrix factorizations such as LU and Cholesky~\cite{conflux, choleskyQRnew}.
% On the other hand, the performance of kernels operating on high-order tensors, for example, the CANDECOMP/PARAFAC (CP)~\cite{cppapers} and Tucker decompositions~\cite{tuckerpaper}, is largely untapped due to the complexity imposed by the high dimensionality of the iteration space.

% \begin{figure}[t]
%     \centering
%     \includegraphics[width=.95\columnwidth]{fig1.pdf}
%     \caption{A runtime comparison between Cyclops Tensor Framework (CTF), the state-of-the-art library for tensor programs, and  Deinsum; we outperform CTF on 95/100 cases.}
%     \label{fig:front-page-figure-placeholder}
% \end{figure}
\begin{figure}[t]
    \centering
    \includegraphics[width=\linewidth]{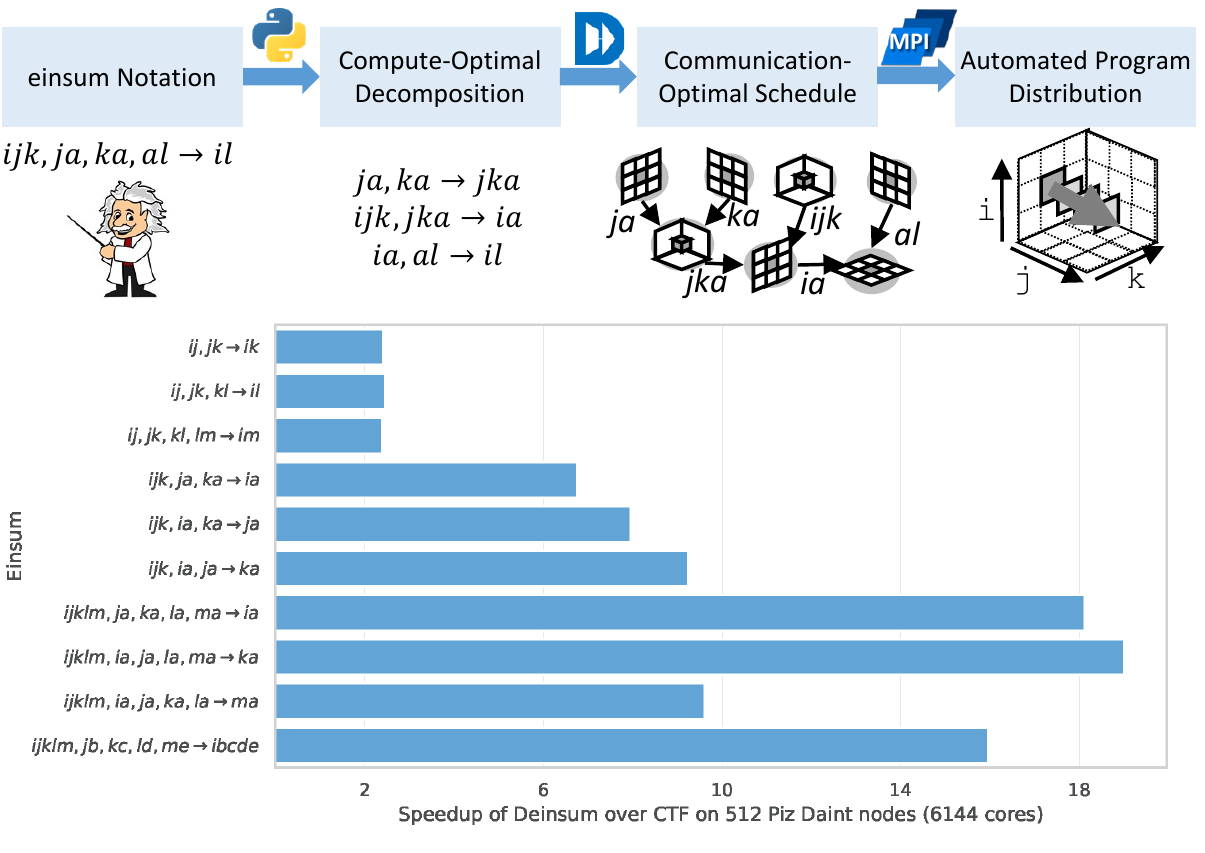}
    \caption{Overview of Deinsum}
    \label{fig:front-page-figure-placeholder}
    \vspace{-2em}
\end{figure}

\begin{figure*}
    \includegraphics[width=\textwidth]{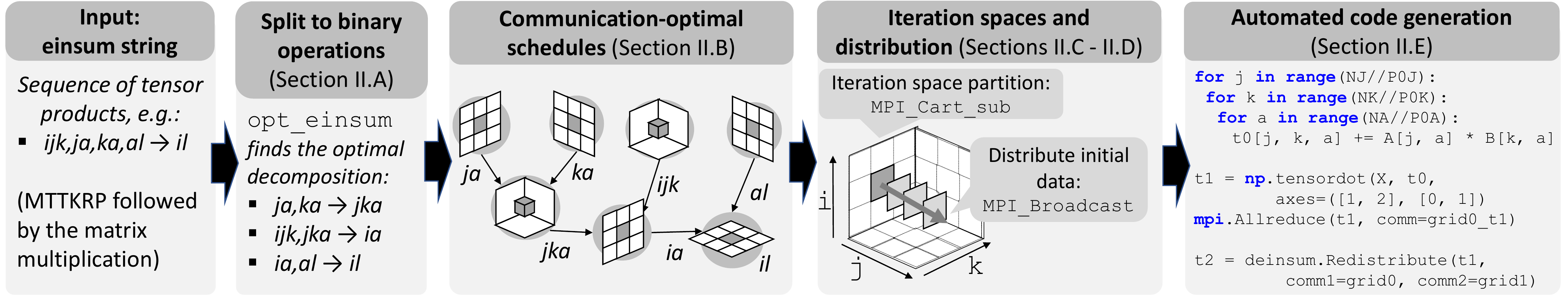}
    \caption{Deinsum accepts arbiratry einsum strings. The single $n$-ary operation is decomposed into a sequence
    of binary operations that minimize the arithmetic complexity. Then, the framework creates the data movement model and automatically derives the 
    tight I/O lower bound together with the corresponding parallel schedule. Next, it creates 
    required iteration spaces, communicators, and data distribution routines. Finally, the entire schedule
    is automatically translated to a high-performance distributed code.}
    \label{fig:workflow}
    \vspace{-1.5em}
\end{figure*}

To close the gap between the well-studied and optimized BLAS- and LAPACK-like kernels and the mostly uncharted data movement modeling in the multilinear territory,  we introduce Deinsum, a framework for the automated derivation of I/O optimal parallel schedules of arbitrary multilinear algebra kernels described in einsum notation \revisionanotag{and operating on dense data.}
To the best of our knowledge, this is the first work that incorporates an analytical model of data reuse and communication minimization 
\revisionanotag{across multiple statements of larger kernels}
% across multiple kernels of large programs
with fully automatic cross-platform code generation, data distribution, and high-performance computation.
The presented pipeline not only provides tight I/O lower bounds for input programs, but also outputs provably 
communication-optimal distributed schedules.
In summary, we make the following contributions:
\begin{itemize}
    % \item Python code-generation framework for fully-automated data distribution and computation on large-scale machines.
    \item Code-generation framework written in Python, fully-automating data distribution and computation at scale.
    \item Tight I/O lower bounds for MTTKRP, the main computational kernel of the CP decomposition, improved by more than $6\times$ over the previously best-known result~\cite{ballard2018communication}.
    \item Up to $19\times$ performance improvement over the current state-of-the-art, Cyclops Tensor Framework (CTF)~\cite{ctf}.
    % , summarized in Fig~\ref{fig:front-page-figure-placeholder}.
\end{itemize}

The rest of the paper is organized as follows.
First, we provide in Sec.~\ref{sec:workflow} a top-down example that introduces Deinsum's workflow together with a high-level description of all the theoretical concepts.
We proceed with a rigorous mathematical formulation of Deinsum, starting with the basic tensor algebra (Sec.~\ref{sec:tensor-defs}), continuing with our framework's underlying I/O lower bound theory (Sec.~\ref{sec:single-bound}), and finishing with the distribution of \revisionanotag{multilinear algebra kernels}
% programs
(Sec.~\ref{sec:distribution}).
Then, we introduce a set of benchmarks that we subsequently use to exhibit Deinsum's superiority against the current state-of-the-art (Sec.~\ref{sec:evaluation}).
% We close the paper with a related work section on the concepts introduced.
We close the paper with a related work section.

\section{Workflow}
\label{sec:workflow}

To describe our framework's workflow, we use as an example the \revisionanotag{multilinear algebra kernel}
%program
described by $ijk,ja,ka,al \rightarrow il$ in \textit{Einstein index notation}.
% This notation is described in more detail in Sec.~\ref{sec:notation}.
% However,
In practical terms, it describes a program with five nested loops, one for each index $\left(i, j, k, a, l\right)$ that appears in the formula.
Each loop iterates over an integer interval, for example, $i \in 0..N_i - 1$, $j \in 0..N_j - 1$, and so on, generating a 5-dimensional \textit{iteration space} equal to the Cartesian product of the five intervals.
We use the notation $\smash[b]{\bm{I} \equiv \bigtimes_{idx \in \left( i, j, k, l, a\right)}\{0..N_{idx}-1\}}$ to describe this space.
The program has four input tensors, one for each of the index strings $ijk$, $ja$, $ka$, and $al$ that appear before the arrow in the formula; an order-3 tensor $\bm{\mathcal{X}}$ with size $N_i N_j N_k$, and three order-2 tensors (matrices) $\bm{A}$, $\bm{B}$, and $\bm{C}$, with sizes $N_j N_a$, $N_k N_a$, and $N_a N_l$ respectively.
The program's output is represented by the index string $il$ that appears after the arrow in the formula and corresponds to a matrix of size $N_i N_l$.
A naive implementation in Python follows:
%\begin{lstlisting}[label=lst:flop-naive]
%for i in range(NI):
%  for j in range(NJ):
%    for k in range(NK):
%      for l in range(NL):
%        for a in range(NA):
%          out[i, l] += (X[i, j, k] * A[j, a] *
%                        B[k, a] * C[a, l])
%\end{lstlisting}
\begin{lstlisting}[label=lst:flop-naive, caption={Naive implementation of $ijk,ja,ka,al \rightarrow il$.}]
for i in range(NI):
 for j in range(NJ):
  for k in range(NK):
   for l in range(NL):
    for a in range(NA):
     out[i,l]+=X[i,j,k]*A[j,a]*B[k,a]*C[a,l]
\end{lstlisting}

The rest of this section provides a high-level overview of Deinsum's inner workings and the workflow's steps, shown in Fig.~\ref{fig:workflow}, using the above program as an example.
Deinsum decomposes the given einsum string into associative binary operations to expose FLOP-reduction opportunities (Sec.~\ref{sec:work_opt_einsum}).
It then lowers the program to a data-centric intermediate representation (Sec.~\ref{sec:work_dace}), facilitating the extraction of the iteration spaces and the I/O lower bound analysis (Sec.~\ref{sec:work_soap}).
It then block-distributes the program's data and computation using a near I/O optimal parallel schedule (Sec.~\ref{sec:work_distribution}).
The final step is the automated code generation and execution on distributed memory machines (Sec.~\ref{sec:work_codegen}).

\subsection{Decomposition of Associative Operations}
\label{sec:work_opt_einsum}
The above implementation is not compute-efficient: there is a significant amount of repeated arithmetic operations since, for every operand,
only a subset of iteration variables is used.
For example, the same multiplication
\texttt{B[k,a] * C[a,l]} is performed for each different value of \texttt{i} and \texttt{j}.
Exploiting the associativity of multiplication, we can break down 
the above $4$-ary operation operation to a series of binary operations, effectively reducing
the overall arithmetic complexity from $4N_iN_jN_kN_lN_a$ to just $2N_iN_a(N_k(1 + N_j) + N_l)$ FLOPs:
\begin{itemize}
    \item $ja,ka \rightarrow jka$ (\texttt{NJ * NA * NK} iterations)
    \item $ijk,jka \rightarrow ia$ (\texttt{NI * NJ * NK * NA} iterations)
    \item $ia,al \rightarrow il$ (\texttt{NI * NA * NL} iterations)
\end{itemize}
The first binary operation is a Khatri-Rao Product (KRP), the second operation is a Tensor Dot Product (TDOT), and the last is matrix multiplication.
All operations that may appear in tensor programs are formally defined in Sec.~\ref{sec:tensor-ops}.
This sequence of operations roughly corresponds to the following Python program that utilizes the NumPy (\texttt{numpy} or \texttt{np})~\cite{numpy} Python module for basic numerical kernels:
\begin{lstlisting}[label=lst:flop-opt]
# ja,ka->jka
t0 = np.zeros((NJ, NK, NA), dtype=X.dtype)
for j in range(NJ):
  for k in range(NK):
    for a in range(NA):
      t0[j, k, a] += A[j, a] * B[k, a]
# ijk,jka->ia
t1 = np.tensordot(X, t0, axes=([1, 2], [0, 1])
# ia,al->il
out = t1 @ C
\end{lstlisting}
Our framework uses \texttt{opt\_einsum}~\cite{opt-einsum} -- the Optimized Einsum Python module, which accepts as input arbitrary \revisionanotag{multilinear algebra kernel} descriptions in Einstein index notation and breaks them down to sequences of binary tensor operations that minimize the overall FLOP count.

% To find a data movement optimal distribution, the first step is to decompose the above 4-ary program to binary operations that operate among the input tensors and intermediate results.
% Our framework uses the Optimized Einsum (\texttt{opt\_einsum})~\cite{opt\_einsum} Python module, which accepts as input arbitrary multilinear algebra program descriptions in Einstein index notation and breaks them down to sequences of tensor binary operations that minimize the overall FLOP count.
% The decomposition suggested by \texttt{opt\_einsum} for the above program is the following sequence:
% \begin{itemize}
%     \item $ja,ka \rightarrow jka$
%     \item $ijk,jka \rightarrow ia$
%     \item $ia,al \rightarrow il$
% \end{itemize}
% The first binary operation is a Khatri-Rao Product (KRP), the second operation is a Tensor Dot Product (TDOT), and the last one is a matrix multiplication.
% All operations that may appear in tensor programs are formally defined in Sec.~\ref{sec:tensor-ops}.
% This sequence of operations roughly corresponds to the following Python program that utilizes the NumPy (\texttt{numpy} or \texttt{np})~\cite{numpy} Python module for basic numerical kernels:
% \begin{lstlisting}
% # ja,ka->jka
% t0 = np.zeros((NJ, NK, NA), dtype=X.dtype)
% for j in range(NJ):
%   for k in range(NK):
%     for a in range(NA):
%       t0[j, k, a] += A[j, a] * B[k, a]
% # ijk,jka->ia
% t1 = np.tensordot(X, t0, axes=([1, 2], [0, 1])
% # ia,al->il
% out = t1 @ C
% \end{lstlisting}

\subsection{Communication-Optimal Parallel Schedules} 
\label{sec:work_dace}
We lower the sequence of binary tensor operations to a data flow-based intermediate representation, utilizing the Data-Centric Parallel Programming (\texttt{dace})~\cite{dace} Python framework.
We extract the program's iteration space and its fine-grained parametric data access patterns from this representation.
We then create a fully-symbolic data movement model, capturing data reuse both inside each binary tensor operation (e.g., via tiling) and across different kernels, potentially reusing intermediate results over the entire chain of operations (e.g., via kernel fusion).
To derive tight I/O lower bounds and corresponding schedules, we implement the combinatorial data access model presented by Kwasniewski et al.~\cite{soap}. 
The outline of the model is presented in Sec.~\ref{sec:single-bound}.
In the above example, our framework outputs the data-movement optimal schedule that fuses the first two binary operations, KRP and TDOT, forming the Matricized Times Tensor Khatri-Rao Product (MTTKRP, defined in Sec.~\ref{sec:tensor-ops}) and then multiplies it with matrix \texttt{C} using a GEMM call with a provided I/O optimal tile size.
We refer to the above two groupings of the program's binary operations as the MTTKRP and MM terms.

Our framework automatically generates one of the theoretical contributions of this work: a tight parallel I/O lower bound for MTTKRP is described in Sec.~\ref{sec:mttkrp-bound}, which provides 
more than $6\times$ improvement over the previously best-known lower bound~\cite{ballard2018communication}.
Interestingly, a two-step MTTKRP (KRP + GEMM), which is commonly used in tensor libraries~\cite{twostep-mttkrp-1,twostep-mttkrp-2}, is not 
communication-optimal (Sec.~\ref{sec:mttkrp-bound}).

\subsection{Iteration Spaces}
\label{sec:work_soap}
For each of the MTTKRP and MM terms we derive their corresponding iteration spaces.
The first term exists in the 4-dimensional space $\smash[b]{\bigtimes_{idx \in \left(i, j, k, a\right)}\{0..N_{idx}-1\}}$, while the second one is in the 3-dimensional space $\smash[b]{\bigtimes_{idx \in \left(i, l, a\right)}\{0..N_{idx}-1\}}$.
The basic idea is to distribute these iteration spaces to the available $P$ processes.
For practicality, we consider distributed programs utilizing MPI communication and, therefore, processes can be assumed to correspond to MPI ranks.
For each iteration space, we arrange the $P$ processes to a Cartesian process grid with the same dimensionality.
The first space is mapped to a grid with dimensions $\smash[b]{(P_i^{(0)}, P_j^{(0)}, P_k^{(0)}, P_a^{(1)})}$, while a $\smash[b]{(P_i^{(1)}, P_a^{(1)}, P_l^{(1)})}$-sized Cartesian grid is generated for the second sub-space.
The superscript of the grid dimensions identifies the term, while the subscript is the dimension index.
The mapping from iteration spaces to Cartesian grids follows the block distribution.
% , which is formally defined in Sec.~\ref{sec:block-distr}.

We note that our framework is parametric in the sizes of the tensors, the optimal tile sizes, and the lengths of the Cartesian process grid dimensions (the number of grid dimensions depends on the dimensionality of the program's iteration space and is constant).
The exact dimensions of the process grids depend on the available number of processes, which can be given at runtime.
To provide a better intuition for the iteration space distributions, we consider for the rest of this section that there are 8 processes and that $N_{idx} = 10$, for $idx \in (i, j, k, l, a)$.
However, a rigorous mathematical model can be found in Sec.~\ref{sec:block-distr}.
According to the tile sizes generated in the previous step, the first grid has dimensions (equivalently, number of tiles) $\left(2, 2, 2, 1\right)$.
This decomposition of the MTTKRP term iteration space to the 8 MPI ranks is shown in Tab.~\ref{tab:space-assignment}.
\begin{table}[h!]
 \scriptsize
 \centering
%   \rowcolors{2}{gray!15}{white}
  \caption{Block distribution of the example program's MTTKRP term iteration space to $P = 8$ MPI processes.}
 \begin{tabular}{ccccc}
  \toprule
  \hiderowcolors
  \multirow{2}{*}[-.12cm]{\textbf{Rank}} & \multicolumn{4}{c}{\textbf{Dimension Slices}}\\
  \cmidrule(l){2-5}
  & $i$ & $j$ & $k$ & $a$\\
  \midrule
  \showrowcolors
  $0$ &   \multirow{4}{*}[-.3cm]{$0..(N_i/2)-1$} & \multirow{2}{*}[-.12cm]{$0..(N_j/2)-1$} & $0..(N_k/2)-1$&
  \multirow{8}{*}[-.6cm]{$0..N_a-1$}\\
  \cmidrule(l){4-4}
  $1$ &   & & $N_k/2..N_k-1$& \\
  \cmidrule(l){3-4}
  $2$ &   & \multirow{2}{*}[-.12cm]{$N_j/2..N_j-1$} & $0..(N_k/2)-1$& \\
  \cmidrule(l){4-4}
  $3$ &   & & $N_k/2..N_k-1$& \\
  \cmidrule(l){2-4}
  $4$ &   \multirow{4}{*}[-.3cm]{$N_i/2..N_i-1$} & \multirow{2}{*}[-.12cm]{$0..(N_j/2)-1$} & $0..(N_k/2)-1$& \\
  \cmidrule(l){4-4}
  $5$ &   & & $N_k/2..N_k-1$& \\
  \cmidrule(l){3-4}
  $6$ &   & \multirow{2}{*}[-.12cm]{$N_j/2..N_j-1$} & $0..(N_k/2)-1$& \\
  \cmidrule(l){4-4}
  $7$ &   & & $N_k/2..N_k-1$& \\
  \bottomrule
  \end{tabular}
  \label{tab:space-assignment}
\end{table}

\subsection{Data and Computation Distribution}
\label{sec:work_distribution}

\begin{figure}
    \includegraphics[width=\linewidth]{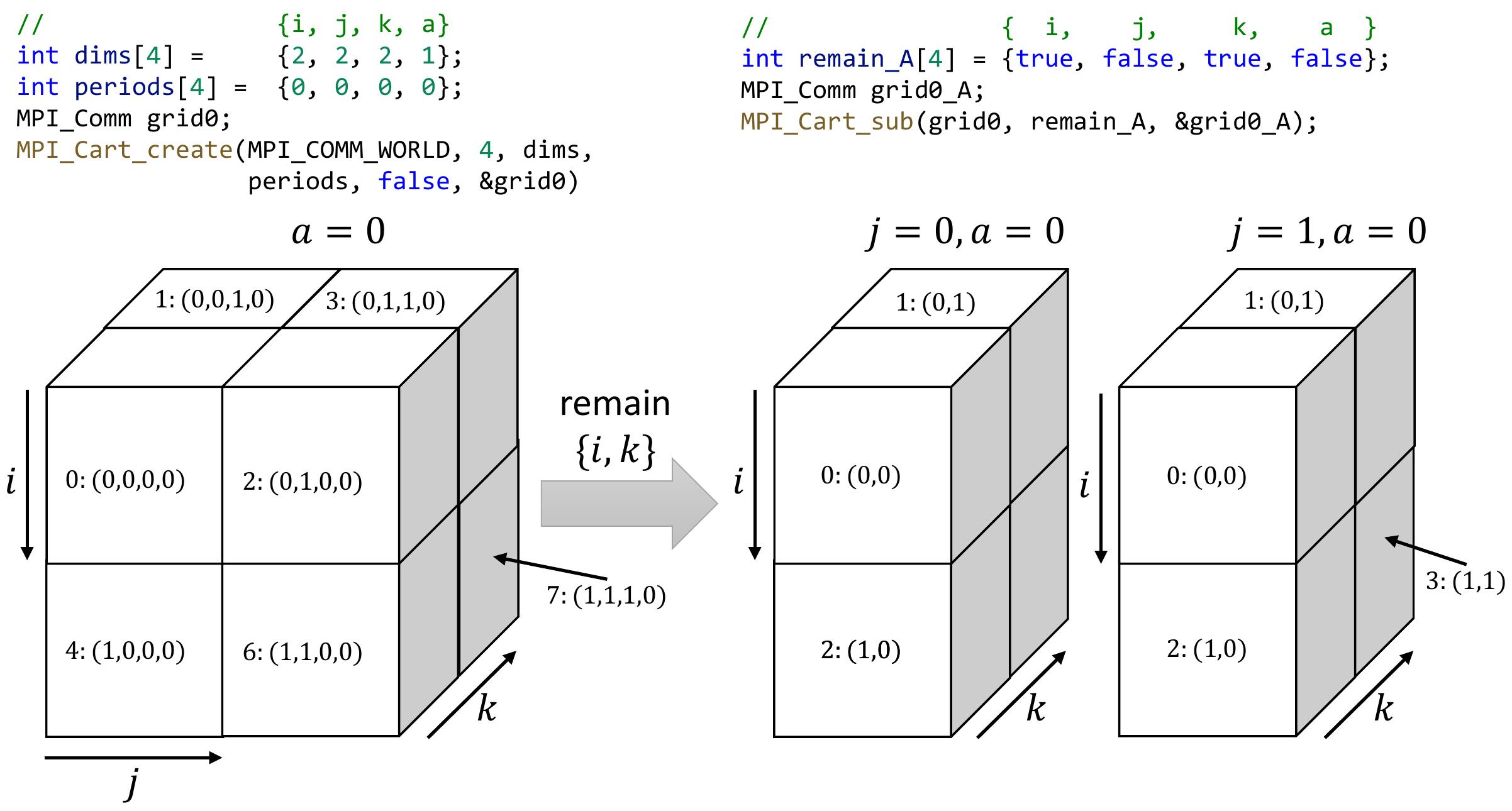}
    \caption{\revisionanotag{The MPI process grids, ranks, and coordinates produced by Listing~\ref{lst:mpi_cart}.}}
    \label{fig:mpi_cart}
    \vspace{-1.5em}
\end{figure}

Subsequently, our framework block-distributes the program's data and computation to the Cartesian process grids straightforwardly; each process is assigned the blocks of data and computation corresponding to its assigned blocks of iteration sub-spaces.
Thus, the input order-3 tensor $\bm{\mathcal{X}}$ is tiled in half in each of its modes, resulting in 8 three-dimensional blocks.
Each process is assigned one of those tiles.
On the other hand, the input matrix $\bm{A}$ is partitioned to only two blocks since the dimension corresponding to the $a$ index is not tiled.
However, if we look again at Tab.~\ref{tab:space-assignment}, we can see that each of these blocks is needed by multiple processes.
For example, both the iteration space blocks assigned to ranks 0, and 1 include the sub-block $\{0..(N_j/2)-1\} \times \{0..N_a-1\}$.
Our framework handles these cases by replicating such data blocks over the necessary processes.
The processes that replicate such data are defined by the Cartesian sub-grids produced by dropping the dimensions that are not relevant to the data.
Readers familiar with MPI Cartesian grids may find it intuitive to consider the \texttt{MPI\_Cart\_create}~\cite{cart-create}, and \texttt{MPI\_Cart\_sub}~\cite{cart-sub} methods.
For example, the Cartesian grid of the first binary operation group and sub-grid for matrix $\bm{A}$ are described by \revisionanotag{the MPI calls in Listing~\ref{lst:mpi_cart}. Furthermore, the result of those calls is visualized in Fig.~\ref{fig:mpi_cart}.}
\begin{lstlisting}[style=cpp, label={lst:mpi_cart}, caption={MPI calls generating the sub-grid for matrix $\bm{A}$}.]
//                {i, j, k, a}
int dims[4] =     {2, 2, 2, 1};
int periods[4] =  {0, 0, 0, 0};
MPI_Comm grid0;
MPI_Cart_create(MPI_COMM_WORLD, 4, dims,
                periods, false, &grid0);
//                {  i,    j,     k,    a  }
int remain_A[4] = {true, false, true, false};
MPI_Comm grid0_A;
MPI_Cart_sub(grid0, remain_A, &grid0_A);
\end{lstlisting}
The \texttt{MPI\_Cart\_sub} call will produce in total $P^{(0)}_j \cdot P^{(0)}_a = 2$ sub-grids, one for each of the $\bm{A}$-blocks.
Each sub-grid includes $P^{(0)}_i \cdot P^{(0)}_k =4$ of the total $P=8$ processes, which replicate one of the $\bm{A}$-blocks.
The assignment of the $\bm{\mathcal{X}}$- and $\bm{A}$-blocks is presented in Tab.~\ref{tab:input-assignment}.
% \alexnick{Discuss shortly the memory issue?}

Block-distributing the data with replication ensures that each process can perform its assigned computation for a specific group of binary operations and iteration sub-space without further inter-process communication, as long as there are no dependencies on intermediate results.
There are two issues to address here.
First, if the output data of the group of operations does not span the whole iteration space, then, in an analogous manner to input replication, each output block is split into partial results.
For example, the output \texttt{t1} of the first group of operations has size $N_iN_a$.
Therefore, there are $P^{(0)}_i \cdot P^{(0)}_a =2$ \texttt{t1}-blocks, each one assigned to $P^{(0)}_j\cdot P^{(0)}_k =4$ MPI ranks.
After completing its assigned computation, each process holds a partial result of its assigned \texttt{t1}-block.
By reducing these partial results over each sub-grid with a collective operation (\texttt{MPI\_Allreduce}), we achieve block distribution with replication for the output so that it can be used in next steps.
The second issue relates to intermediate data that are the output of one group of operations and the input to another.
In general, the distributions of the groups differ, and the data must be redistributed.
In our 8 process example, the second group of operations has an iteration sub-space $\smash[b]{\bigtimes_{idx \in \left(i, l, a\right)}\{0..N_{idx}-1\}}$, assigned to a process grid with sizes $\left(2, 2, 2\right)$.
The intermediate tensor \texttt{t1} must be redistributed from a block distribution over $P^{(0)}_i \cdot P^{(0)}_a =2$ processes to another block distribution over $P^{(1)}_i \cdot P^{(1)}_a = 4$ processes.
Our framework automatically infers the communication needed to redistribute tensors across different block distributions and Cartesian process grids.
The redistribution's theoretical background is presented in Sec.~\ref{sec:redistr}.

\subsection{Automated Code Generation}
\label{sec:work_codegen}
The last step involves putting all the above analyses together and generating code that executes \revisionanotag{multilinear algebra kernels} in distributed machines using MPI.
To that end, we employ again \texttt{dace}, which includes basic MPI support~\cite{P4SC21}.
We extend this functionality to support MPI Cartesian grids, and we use the available API to implement our redistribution scheme as a library call.
We create the distributed program by adding the necessary MPI communication calls to the intermediate representation.
\revisionanotag{The generated distributed code (for a specific einsum) is then compiled to a shared library that can be called by any application.
Furthermore, Deinsum outputs an intermediate Python program that is functionally equivalent to the generated code.
For the example presented in this section, the Python code is the following:}
% The generated distributed code is functionally equivalent to the following Python code:
\begin{lstlisting}
grid0 = mpi.Cart_create(dims=[P0I,P0J,P0K,P0A])
grid0_t1 = mpi.Cart_sub(
  comm=grid0, remain=[False,True,True,False])
grid1 = mpi.Cart_create(dims=[P1I, P1L, P1A])
grid1_out = mpi.Cart_sub(
  comm=grid1, remain=[False,False,True])
# ja,ka->jka
t0 = np.zeros((NJ//P0J, NK//P0K, NA//P0A),
              dtype=X.dtype)
for j in range(NJ//P0J):
  for k in range(NK//P0K):
    for a in range(NA//P0A):
      t0[j, k, a] += A[j, a] * B[k, a]
# ijk,jka->ia
t1 = np.tensordot(X, t0, axes=([1, 2], [0, 1])
mpi.Allreduce(t1, comm=grid0_t1)
# ia,al->il
t2 = deinsum.Redistribute(t1, comm1=grid0,
                              comm2=grid1)
out = t2 @ C
mpi.Allreduce(out, comm=grid1_out)
\end{lstlisting}

\begin{table}[h!]
 \footnotesize
 \centering
 \caption{Block-distribution with replication of the example program's tensors $\bm{\mathcal{X}}$, and $\bm{A}$ to $P = 8$ MPI processes, with $N_{idx} = 10$.}
  \rowcolors{2}{gray!15}{white}
 \begin{tabular}{p{.64cm}p{1.24cm}p{3.55cm}p{1.75cm}}
  \toprule
  \textbf{Rank} & \textbf{Coords} & $\bm{\mathcal{X}}$-\textbf{Block} & $\bm{A}$-\textbf{Block}\\
  \midrule
  $0$ &   $\left(0, 0, 0, 0\right)$ &  \lstinline[style=tablestyle]$X[:5, :5, :5]$   &   \lstinline[style=tablestyle]$A[:5, :]$\\
  $1$ &   $\left(0, 0, 1, 0\right)$ &  \lstinline[style=tablestyle]$X[:5, :5, 5:]$   &   \lstinline[style=tablestyle]$A[:5, :]$\\
  $2$ &   $\left(0, 1, 0, 0\right)$ &  \lstinline[style=tablestyle]$X[:5, 5:, :5]$   &   \lstinline[style=tablestyle]$A[5:, :]$\\
  $3$ &   $\left(0, 1, 1, 0\right)$ &  \lstinline[style=tablestyle]$X[:5, 5:, 5:]$   &   \lstinline[style=tablestyle]$A[5:, :]$\\
  $4$ &   $\left(1, 0, 0, 0\right)$ &  \lstinline[style=tablestyle]$X[5:, :5, :5]$   &   \lstinline[style=tablestyle]$A[:5, :]$\\
  $5$ &   $\left(1, 0, 1, 0\right)$ &  \lstinline[style=tablestyle]$X[5:, :5, 5:]$   &   \lstinline[style=tablestyle]$A[:5, :]$\\
  $6$ &   $\left(1, 1, 0, 0\right)$ &  \lstinline[style=tablestyle]$X[5:, 5:, :5]$   &   \lstinline[style=tablestyle]$A[5:, :]$\\
  $7$ &   $\left(1, 1, 1, 0\right)$ &  \lstinline[style=tablestyle]$X[5:, 5:, 5:]$   &   \lstinline[style=tablestyle]$A[5:, :]$\\
  \bottomrule
  \end{tabular}
  \label{tab:input-assignment}
\end{table}

\section{Tensor Algebra}

This section describes the mathematical notation and the fundamental concepts behind data movement analysis in multilinear algebra.
We use $0$-based indexing for consistency.

\subsection{Tensor Definitions and Einstein Summation Notation}
\label{sec:tensor-defs}

\revisionanotag{Multilinear algebra programs operate on tensors, frequently represented by multidimensional arrays. The formal definition
of tensors, tensor spaces, and their mathematical significance as basis-independent transformations is beyond the scope of this paper --- rigorous definitions can be found in dedicated literature~\cite{tensorBook}. In this work, we focus on tensors from the computational and data movement
perspectives, thus we refer to 
an order $N$ tensor $\bm{\mathcal{X}}$ simply as an element of an $N$ dimensional vector space $\bm{\mathcal{X}} \in \mathcal{F}^{I_0\times ...I_{N-1}}$
over field $\mathcal{F}$, where $\mathcal{F}$ is typically the field of real $\mathbb{R}$ or complex numbers $\mathbb{C}$. Analogously, we refer to vectors as order 1 tensors and to matrices as order 2 tensors.}
%
% An $N$-dimensional array $\bm{\mathcal{X}} \in \mathcal{F}^{I_0\times ...I_{N-1}}$ is a tensor of \textit{order} $N$ or, equivalently, a tensor with $N$ \textit{modes}.
% Consequently, a tensor of order 1 is a vector, while matrices are tensors of order 2.
% The tensor is defined in the $N$-dimensional vector space $\mathcal{F}^{I_0\times \dots I_{N-1}}$, where $\mathcal{F}$ is the field of the tensor values, typically the real $\mathbb{R}$, or complex numbers $\mathbb{C}$.
%
We define $\smash[b]{\bm{I} = \bigtimes_{j\in 0..N-1}{I_j}}$ as the tensor's \textit{iteration space}. The set of indices $(i_0, i_1, \dots, i_{N-1})$ that iterate over 
$\smash[b]{\bm{I}}$ is used to access tensor elements.
Given a multilinear map $f$

$$f: V_0 \times \dots \times V_{N-1} \rightarrow W$$

where $V_0, \dots, V_{N-1}$, and $W$ are vector spaces, while $f$ is a linear function w.r.t. each of its $N$ arguments,
this map has the associated tensor product:

$$ \bm{\mathcal{W}} = \bm{\mathcal{V}}^0 \otimes \dots \otimes \bm{\mathcal{V}}^{N-1}$$

where $\bm{\mathcal{W}} \in W$, and $\bm{\mathcal{V}}^j \in V_j$. $\bm{\mathcal{V}}^j$ are tensor \emph{modes}.
Assuming that the tensors have iteration spaces $\bm{I}^w$, and $\bm{I}^{j}$ respectively, the above expression can be simplified using the Einstein summation notation:
$$\mathcal{W} = \mathcal{V}^0_{I_0} \mathcal{V}^1_{I_1} \dots \mathcal{V}^{N-1}_{I_{N-1}}$$
Repeated indices in the iteration spaces of the $\bm{\mathcal{V}}^j$ tensors are implicitly summed over, while non-repeated indices correspond to dimensions of $\bm{\mathcal{W}}$. 

To provide an intuitive example, $y = A_{j i} A_{i k} x_{k}$
represents the equation $\bm{y} = \bm{A}^T\cdot\bm{A}\cdot\bm{x}$, where the repeated indices $i$ and $k$ represent reduction over corresponding dimensions and the final result is a one-dimensional vector with index $j$. Analogously, $C = A_{ik}B_{kj}$ is the matrix-matrix product, and $A = u_i v_j$ is the outer product of vectors $\bm{u}$ and $\bm{v}$.
We note that it is common, especially in programming libraries that implement the Einstein notation, to drop the tensor names and keep only the indices. For example, the $y = A_{j i} A_{i k} x_{k}$ expression is simplified to \texttt{ji,ik,k->j}.
The three index string before the right arrow are the \textit{access indices} of the three input tensor, while \texttt{j} is the access index of the output $\bm{y}$.

\subsection{Tensor operations}
\label{sec:tensor-ops}

Having defined tensors, we proceed with describing basic tensor operations that frequently appear in \revisionanotag{multilinear algebra kernels.}
For the rest of this section we use the tensor $\bm{\mathcal{X}} \in \mathcal{F}^{I_0\times ...I_{N-1}} \equiv X$.
We start with a unary tensor operation,  the \textit{mode-$n$ matricization} $\bm{\mathcal{Y}} = \bm{\mathcal{X}}_{(n)}$:
\begin{align}
    f:\hspace{.5em} X &\rightarrow \mathcal{F}^{(I_0\dots I_{n-1}I_{n+1}\dots I_{N-1})\times I_n}, \quad\bm{\mathcal{X}} \mapsto \bm{\mathcal{X}}_{(n)}\nonumber
\end{align}

In simple terms, this operation transposes a tensor by permuting the order of its modes so that the $n$-th mode comes last (or first, depending on convention).
Subsequently, it \textit{flattens} the first (or last) $N-1$ modes, effectively transforming the tensor to a matrix.
The flattening of the modes cannot be expressed as an einsum, however, the transposition can be written as $i_0\dots i_{N-1}\rightarrow i_0\dots i_{n-1}i_{n+1}\dots i_{N-1}i_n$.
Next is the \textit{mode-$n$ tensor product} or Tensor Times Matrix (TTM), denoted by $\times_{n}$. It is an operation in mode-$n$ between a tensor $\bm{\mathcal{X}}$ and a matrix $\bm{U} \in \mathcal{F}^{I_n\times R}\equiv U$:

{\footnotesize
\begin{align}
    f:\hspace{.5em} X \times U&\rightarrow \mathcal{F}^{I_0\times \dots I_{n-1}\times R\times I_{n+1}\times \dots I_{N-1}}, \quad\left(\bm{\mathcal{X}}, \bm{U}\right) \mapsto \bm{\mathcal{X}}\times_{n}\bm{U}\nonumber
\end{align}
}%
This product is computed by multiplying each of the tensor's mode-$n$ vectors (\textit{fibers}) by the $\bm{U}$ matrix.
Another way to compute TTM is to produce the mode-$n$ matricization of the $\bm{\mathcal{X}}$ tensor, multiply by the matrix $\bm{U}$ and $\textit{fold}$ the output matrix back to an order-$N$ tensor so that the fibers corresponding to $\bm{U}$'s columns are placed in the $n$-th mode.
This operation's einsum is $i_0\dots i_{N-1},i_nr\rightarrow i_0\dots i_{n-1}ri_{n+1}\dots i_{N-1}$.
The Khatri-Rao Product (KRP) is defined as the column-wise Kronecker product of two matrices:
\begin{align}
    f:\hspace{.5em} \mathcal{F}^{I_0\times R}\times\mathcal{F}^{I_1\times R} &\rightarrow \mathcal{F}^{I_0\times I_1}, \quad\left(\bm{U}^{0}, \bm{U}^{1}\right) \mapsto \bm{U}^{0}\odot\bm{U}^{1}\nonumber
\end{align}
Its einsum representation is $i_0r,i_1r\rightarrow i_0i_1$.
We note that both TTM and KRP can operate on tensors that are matricized appropriately, allowing TTM to generalize to the Tensor Dot Product (TDOT).
Multiple TTM operations can be chained together to form a mode-$n$ Tensor Times Matrix chain (TTMc).
This is an $(N-1)$-ary operation on an order-$N$ tensor and $N-1$ matrices $\bm{U}^{j} \in \mathcal{F}^{I_j\times R_j}\equiv U^j, j \in \{0..N-1\}\setminus{n}$:
\begin{align}
    f&:\hspace{.5em} X\times U^0 \times \dots U^{n-1}\times U^{n+1} \times \dots U^{N-1}\nonumber\\
    &\rightarrow \mathcal{F}^{R_0\times \dots R_{n-1}\times I_n\times R_{n+1}\times \dots R_{N-1}}\nonumber\\
    &\left(\bm{\mathcal{X}}, \bm{U}^{0},\dots,\bm{U}^{n-1},\bm{U}^{n+1},\dots,\bm{U}^{N-1}\right)\nonumber\\
    &\mapsto \bm{\mathcal{X}}\times_{0}\bm{U}^{0}\dots\times_{n-1}\bm{U}^{n-1}\times_{n+1}\bm{U}^{n+1}\dots\times_{N-1}\bm{U}^{N-1}\nonumber
\end{align}
TTMc is written as the einsum:
\begin{align}
    &i_0\dots i_{n-1}i_{n+1}\dots i_{N-1},i_0r_0,\dots,i_{n-1}r_{n-1},\nonumber\\
    &i_{n+1}r_{n+1},\dots,i_{N-1}r_{N-1} \rightarrow r_0\dots r_{n-1}i_nr_{n+1}\dots r_{N-1}\nonumber
\end{align}
The mode-$n$ Matricized Tensor Times Khatri-Rao Product (MTTKRP) is defined in a similar manner:
\begin{align}
    f&:\hspace{.5em} X\times U^0 \times \dots U^{n-1}\times U^{n+1} \times \dots U^{N-1}\rightarrow \mathcal{F}^{I_n\times R}\nonumber\\
    &\left(\bm{\mathcal{X}}, \bm{U}^{0},\dots,\bm{U}^{n-1},\bm{U}^{n+1},\dots,\bm{U}^{N-1}\right)\nonumber\\
    &\mapsto \bm{\mathcal{X}}\odot\bm{U}^{0}\odot\bm{U}^{n-1}\odot\bm{U}^{n+1}\dots\odot\bm{U}^{N-1}\nonumber
\end{align}
It is described in Einstein notation by
{\small
\begin{align}
    i_0\dots i_{n-1}i_{n+1}\dots i_{N-1},i_0r,\dots,i_{n-1}r,i_{n+1}r,\dots,i_{N-1}r\rightarrow i_nr\nonumber
\end{align}
}%

%  \subsection{Iteration Spaces of Multilinear Algebra Programs}
%  \label{sec:spaces}

% Programmatically, the evaluation of an einsum may be expressed as nested \texttt{for} loops of depth equal to the number of unique indices that are used to address the arrays. For example, the $\bm{y} = \bm{A}^T\bm{A}\bm{x}$ product presented in Sec.~\ref{sec:tensor-defs} has the following naive implementation:

% \begin{lstlisting}[caption={An implementation of $y = A_{i j} A_{j k} x_{k}$ as a single-step computation.}, label={lst:atax}]
% for i in range(N):
%   for j in range(N):
%      for k in range(N):
%         y[i] += a[i,j] * a[j,k] * x[k]
% \end{lstlisting}

% \alexnick{wooooosh, greg you talk about iteration space reduction. We need to somehow make the slightly different iteration space definitions agreeable.}

% This exposes two paths for optimizations:
% \begin{itemize}
%     \item taking advantage of associativity to reduce the iteration space size: $(A \cdot A) \cdot x$ vs $A \cdot (A \cdot x)$, 
%     which reduces the iteration space size from $\mathcal{O}(N^3)$ to $\mathcal{O}(N^2)$
%     \item data reuse, since every element of tensor $v$ is accessed as many times as the product of dimensions of indices not used by $v$
% \end{itemize}

\section{Tight data movement lower bounds for multilinear algebra kernels}
\label{sec:single-bound}
% Observe that each execution of a nested loop (see, for example, Listing~\ref{lst:flop-naive}) has an associated \textit{iteration vector}
% $\bm{\psi}_t = [\psi_t^1, \dots, \psi_t^d], t = 1, \dots, |V|$, where $V \in \mathbb{Z}^d$ represents the $d$-dimensional 
% iteration space (e.g., all $N^3$ points in a 3D iteration space of matrix-matrix multiplication).
\revisionanotag{
We now introduce our data movement model of multilinear algebra kernels. 
As discussed in Section~\ref{sec:workflow}, the evaluation of such programs may be encoded as an scalar addition-multiplication in an $n$-deep loop nest (Listing~\ref{lst:flop-naive}). Each execution of this
operation has an associated \emph{iteration vector} $\bm{\psi} = $\texttt{[i, j, k, l, a]} of iteration variables' values. The central idea behind finding the data movement lower bounds is to bound the minimum
number of tensor elements that needs to be loaded/communicated to perform a given number of elementary operations.}

\subsection{Data reuse and computational intensity}
Consider an arbitrary sequence of $X$ elementary operations and their associated iteration vectors $\bm{\Psi} = \{\bm{\psi}_{t_0}, \dots, \bm{\psi}_{t_1}\}$, with 
$|\bm{\Psi}| = t_1 - t_0 = X$. Equivalently, $\bm{\Psi}$ is a set of $X$ new computed values. However, the evaluation of $\bm{\Psi}$ may 
require $Q(\bm{\Psi}) < |\bm{\Psi}|$ I/O operations from main memory, since some elements may be reused while residing in fast memory. 
For example, in classical matrix multiplication \texttt{C[i,j]+=A[i,k]*B[k,j]}, for each different value of the iteration variable
\texttt{j}, the previously loaded element \texttt{A[i,k]} is \emph{reused}. It has been proven~\cite{cosma} that to perform
any execution set $\bm{\Psi}$ of this kernel, with $|\bm{\Psi}| = X$ on a machine with fast memory of size $S$,
at least $Q(\bm{\Psi}) \ge \frac{2X}{\sqrt{S}}$ elements have to be loaded to the fast memory. Equivalently, 
$\rho = \frac{\sqrt{S}}{2}$ is the \emph{computational intensity} of this kernel. Intuitively, for each loaded element, no more than $\rho$ new elements can be computed.

\subsection{Automated derivation of data movement lower bounds}
Kwasniewski et al.~\cite{soap} defined a class of programs called SOAP - Simple Overlap Access Programs. We refer readers to the original paper
for the full formal definition, but for our purposes, it suffices to observe that all multilinear algebra kernels considered in this paper,
such as tensor contractions and decompositions, belong to the SOAP class. The authors further derived a proof of data movement lower bounds
for programs that belong to this class. Below we present a summary of the four main lemmas.

\begin{lma}[Intuition behind Lemmas 1-4~\cite{soap}]
Total data movement volume from the main memory to the fast memory of size $S$ of a program that computes array $A_0$ as a function of input arrays $A_1, \dots, A_n$ 
inside a nested loop is bounded by 
$$Q \ge \frac{|V|}{\rho},$$
% where $|V|$ is the size of the iteration space of the nested loop and $\rho$ is the computational intensity.
where $|V|$ is the nested loop's iteration space size and $\rho$ is the computational intensity.
$\rho$ can be 
further bounded by
$$\rho \le \max_{\bm{\Psi}} \frac{(\sum_{i=1}^n|\mathcal{A}_i(\bm{\Psi})|) - S}{|\bm{\Psi}|},$$
where, for any given set of executions $\bm{\Psi}$, $\mathcal{A}_i$ is a set of elements of input array $A_i$ accessed during $\bm{\Psi}$.
\end{lma}

\subsection{Programs containing multiple statements}
\label{sec:multi-bound}

In multilinear algebra, analyzed problems often require contracting or decomposing multiple tensors with many intermediate values.
Due to the associativity of the addition and multiplication operations, 
a program expressed in the Einstein notation as $w = v^1_{I_1} v^2_{I_2} \dots v^n_{I_n}$ may be written as a sequence of $n-1$ binary operations
$w^0_{W_0} = v^1_{I_1} v^2_{I_2}$, $w^1_{W_1} = w^0_{W_0} v^2_{I_2},$ $\dots,$ $w^n_{W_n} = w^{n-1}_{W_{n-1}} v^n_{I_n}$.
%, with $w^0, \dots, w^{n-1}$ being partial results. Observe that this can asymptotically reduce the iteration space (see Listing~\ref{lst:flop-opt}).
Observe that this can asymptotically reduce the iteration space (see Listing in Sec.~\ref{sec:work_opt_einsum}).
Finding the order of contraction that minimizes the total number of arithmetic operations is NP-hard in a general case~\cite{tensor-order-nphard}.
However, it is possible to exhaustively enumerate all combinations for a small enough number of tensors and select the optimal one.

While the arithmetic complexity of programs containing multiple statements is easy to analyze --- the arithmetic complexity of 
a program is the sum of complexities of each constituent statement-- this is not the case for the I/O complexity.
Data reuse between multiple statements (e.g., caching intermediate results or fusing computations that share the same inputs)
can asymptotically reduce the overall I/O cost. Loop fusion is one of the examples of this problem and is proven to be
NP-hard~\cite{loopFusionComplexity}. However, analogous to the optimal contraction permutation problem, for small enough 
problems, Kwasniewski et al. ~\cite{soap} designed an abstraction that can precisely model data reuse between statements, 
proving the I/O lower bound for programs containing multiple statements. The key component of the method is the 
Symbolic Directed Graph (SDG) abstraction, in which every vertex is a tensor (input or intermediate), and edges represent
data dependencies. Then, each subgraph of non-input SDG vertices represents one of the possible kernel fusions - vertices in the subgraph correspond
to the fused kernels. Each subgraph (and its corresponding fused kernel) can be expressed as a SOAP statement, and its I/O
lower bound is evaluated. By enumerating all possible SDG partitions, the one that minimizes the total I/O cost is chosen and represents
the I/O lower bound of the entire program.

\subsection{Sparse data structures}
\revisionanotag{The data movement model assumes that all data structures 
are dense --- 
each element of input arrays $A_1, \dots, A_n$ is non-trivial and has to be loaded to the fast memory at least once. This assumption is necessary to associate a set of computations  $\bm{\Phi} = \{\bm{\phi_{t_0}}, \dots, \bm{\phi_{t_1}}\}$ with well-defined sets of required input elements $\mathcal{A}_1(\bm{\Phi}), \dots, \mathcal{A}_n(\bm{\Phi})$. However, the model can be extended to sparse data structures using probabilistic methods. Given a probability distribution of non-zero elements in the input tensors $P(A[\bm{\phi}] \ne 0)$, one can derive the expected number of non-zero elements in the access sets $E[|\mathcal{A}_j(\bm{\Phi})|]$. Then, the achieved lower bounds will also be probabilistic and can still be similarly used to obtain data movement-minimizing tilings and data distributions. However, the formal derivation of this extension is beyond the scope of this paper.} 

\subsection{Tight MTTKRP parallel I/O lower bound}
\label{sec:mttkrp-bound}
We now proceed to one of our main theoretical contributions: the MTTKRP I/O lower bound. Contrary to the state-of-the-art approaches,
we show that GEMM-like parallel decomposition is communication-suboptimal. Instead, our new tiling scheme asymptotically reduces the
communication by the factor of $S^{1/6}$, where $S$ is the size of fast local memory. It also improves the previously best-known
lower bound by a factor of $3^{5/3} \approx 6.24$ times~\cite{ballard2018communication}.
\begin{figure}
    \includegraphics[width=0.5\textwidth]{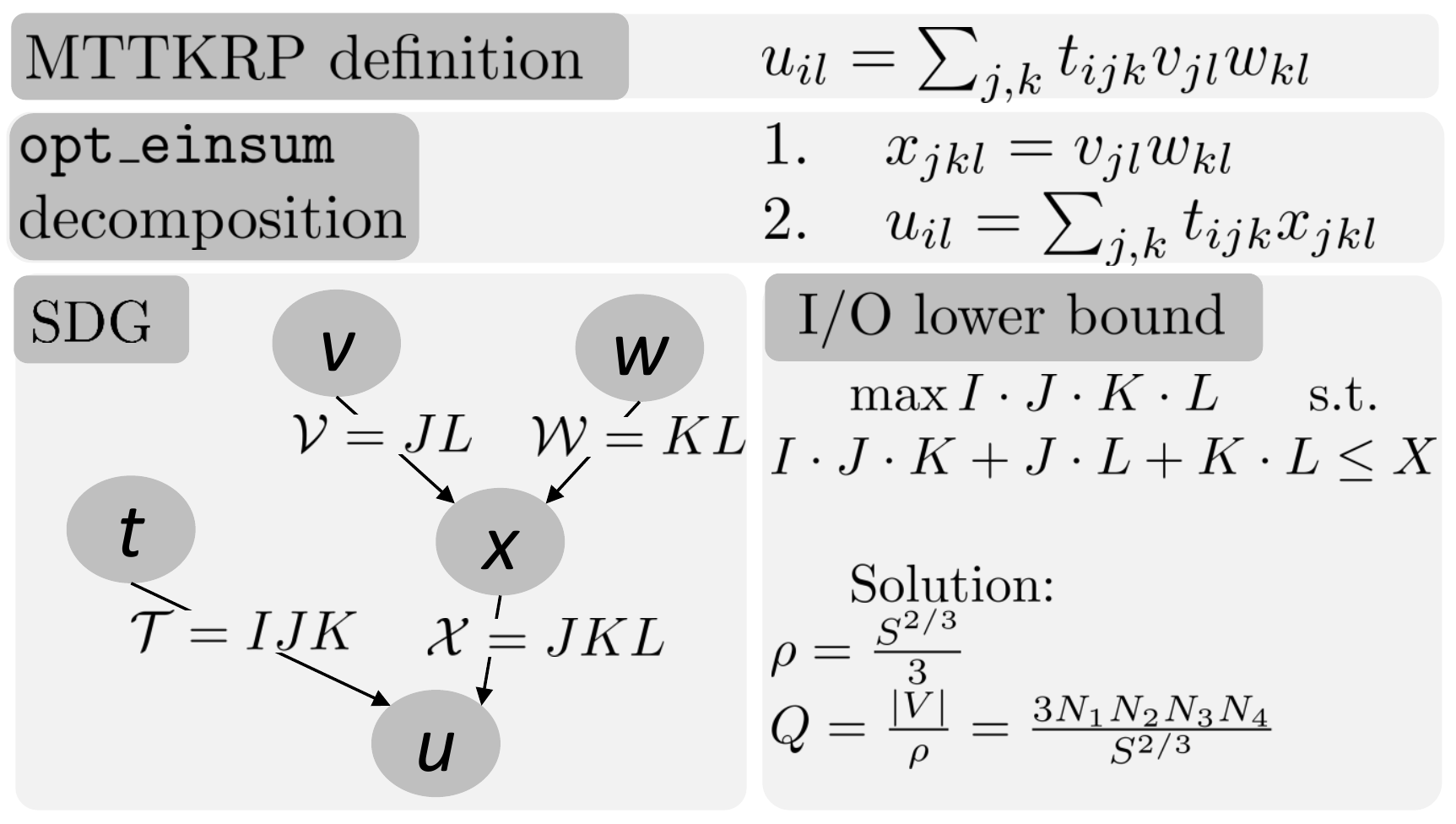}
    \caption{MTTKRP formulation, its breakdown to two operations by the opt\_einsum library, the SDG, and its I/O lower bound. $\mathcal{V}$,
    $\mathcal{W}$, $\mathcal{T}$, $\mathcal{X}$ are the minimum number of elements accessed from tensors $v$, $w$, $T$, and $X$ during a computation that 
    computes $IJKL$ partial products of the output tensor $u$.}
    \label{fig:mttkrp_sdg}
    \vspace{-2em}
\end{figure}

The MTTKRP SDG is shown in Fig.~\ref{fig:mttkrp_sdg}. Observe that there are two possible partitions: $P_1 = \{\{x\}, \{u\} \}$
and $P_2 = \{\{x, u\} \}$. $P_1$ corresponds to a schedule in which
the entire intermediate tensor $x$ is computed first,
and the output tensor $u$ is evaluated next. $P_2$ corresponds to a schedule
when these kernels are fused together and 
every partial result of $x$ is immediately reused to update $u$. 
Consider an arbitrary set of computations $\bm{\Psi}, |\bm{\Psi}| = X$. Denote $I$ the number of different
values iteration variable $i$ takes during $\bm{\Psi}$. Analogously, denote $J$, $K$, $L$ the number of different
values of iteration variables $j$, $k$, $l$. We need to express $I$, $J$, $K$, $L$ as functions of the computation 
size $X$. They represent optimal tile sizes in each of the dimensions that maximize the data reuse.
We now formulate the SOAP optimization problem for $P_2$\cite{soap}:
\begin{align}
\nonumber
    \max I \cdot J \cdot K \cdot L \text{\hspace{2em} s.t} \\
    \nonumber
I\cdot J \cdot K + J \cdot L+ K \cdot L \le X 
\end{align}
which yields
$I(X) = J(X) = K(X) = \sqrt[3]{\frac{2}{5}X}$, \hspace{1em} $L(X) = \frac{X^{2/3}}{\sqrt[3]{2}5^{2/3}}$. \revisionanotag{The interpretation is the following:} for any computation $\bm{\Psi}, |\bm{\Psi}| = X$,
these are the tile sizes that minimize the number of I/O operations. But since this is true for \emph{any} $X$,
and we want to find a tight I/O lower bound, we find the $X_0$ that \emph{maximizes} the I/O cost:
$$X_0 = \text{argmin}_X \frac{I(X)\cdot J(X) \cdot K(X) \cdot L(X)}{X -S}$$
which yields $X_0 = 5S / 2$. Now, substituting $X_0$ to the tile sizes $I(X), J(X), K(X), L(X)$ \revisionanotag{we obtain the final
I/O lower bound, and the corresponding optimal tiling}:
$$\rho = \frac{S^{2/3}}{3} \text{, \hspace{3em}} Q_{MTTKRP} \ge \frac{|V|}{\rho} = \frac{3N_1 N_2 N_3 N_4}{S^{2/3}}$$
$$I = J = K = S^{1/3}, \text{\hspace{2em}} L = S^{2/3} / 2$$
This result not only constitutes a tight I/O lower bound but also provides the corresponding tiling scheme and 
communication-optimal parallel decomposition for any size of the local memory $S$.

% \subsection{MPI stuff}

% \begin{itemize}
%     \item Cartesian topologies
%     \item Block distributions mapped to those topologies
%     \item Redistributions of data among different topologies
% \end{itemize}
% odot
% \subsection{Iteration and Data Spaces}

% \alexnick{In this work, we don't have separate iteration and data spaces, but a single unified space. Shall we make this distinction and explain why (for multilinear algebra) the "optimal" way is to use a unified space or just pretend that the separation concept doesn't exist?}

% In our framework, data are $D$-mode tensors $\mathcal{X} \in \mathbb{C}^{I_1 \times I_2 \times ... \times I_D}$. Without loss of generality, we describe the $j$-mode integer set $I_j$ as consisting of the contiguous integer space $\{0..S_j - 1\}$, where $S_j$ is the tensor's length in the $j$-th dimension. \alexnick{Maybe we should start using "cartesian" here.} multilinear algebra programs generally consist of sets of nested loops that access the tensor data through $D$-dimensional index vectors $\bm{i} = (i_1, i_2, ..., i_D)$, defining their own iteration spaces. However, since these spaces are a subset of the corresponding data spaces, we consider that each program defines a unified data/iteration space $\smash[b]{\bm{I} = \bigtimes_{j=0}^{D-1}\{0..S_j - 1\}}$.

% \alexnick{Here we need to give more details. E.g., the unified data/iteration space is the "union" of all the individual data spaces. Moreover, we should discuss about how we break multilinear algebra programs to individual SDG subgraphs (terminology?) with their own space.}

\section{Distribution of Multilinear Algebra Kernels}
\label{sec:distribution}

% In this work, we consider block distributions of data and computations, which are defined in the next section. To assist with the distribution of these blocks, we assign to the available compute/memory resources (read: MPI processes/ranks) coordinates in multi-dimensional process grids.

% \alexnick{Each multilinear algebra program is decomposed through our framework to SDG subgraphs (what is the correct terminology? @Greg) Each of this SDG subgraphs has its own process grid. We connect two consecutive subgraphs with "redistributions of data".}
This section defines a mathematical framework for describing iteration space distributions. We proceed by defining the block distribution of a \revisionanotag{multilinear algebra kernel's} iteration space on a Cartesian process grid (notation summarized in Tab.~\ref{tab:symbols}).
We then describe the redistribution of data among different block distributions.

\subsection{Iteration Space Distribution}
\label{sec:generic-distr}

Let $\bm{I} = \smash[b]{\bigtimes_{j=0}^{N-1}\{0...I_j-1\}}$ be the $N$-dimensional iteration space of a mulitlinear algebra program, where $I_j$ is the size of the $j$-th dimension.
To \emph{distribute} $I$, we first partition it to any number of disjoint subsets, which consist of consecutive elements in any dimension.
This means that we can uniquely identify each subset by selecting the element with space coordinates $\smash[b]{\bm{b} = \left(b_0, b_1, ..., b_{D-1}\right)^T}$, which has the shortest Euclidean distance from the origin. Any other element in the subset can be defined with respect to $\bm{b}$.
The size of these subsets is defined by a vector $\bm{B} = (B_0, B_1, ..., B_{D-1})^T$.
The subsets may have different sizes, so each vector component $B_j$ may be either a constant or vary depending on $\bm{b}$ or other parameters.
The coordinates $\bm{i}$ of each element in the space can be rewritten as:\vspace{-0.5em}
\begin{align}
    \label{eq:param_dist_idx}
    \bm{i} = \bm{b} + \bm{o}\text{, }
\end{align}
where $\smash{\bm{o} = \left(o_0, o_1, ..., o_{D-1}\right)^T}$, with $o_j \in \{0 ... B_j - 1\}$, are the offset coordinates of the element relative to $\bm{b}$.
We define the one-to-one mappings from the coordinates $\bm{i}$ of an element to the subset it belongs to and to its offset:
\begin{align}
    \label{eq:param_dist_b}
    \bm{b} = u(\bm{i})\\
% \end{align}
% \begin{align}
    \label{eq:param_dist_o}
    \bm{o} = v(\bm{i})
\end{align}
After partitioning the space, we assign the subsets to $P$ processes, with each process having a unique identifier $\bm{p}$.
Since each process may be assigned multiple subsets, we define a second unique subset identifier $\bm{l}$, which is process-local.
We do not set any requirements in the form of $\bm{p}$ and $\bm{l}$, which may be, e.g., scalars or vectors, depending on the distribution.
We define mappings among the global subset identifier $\bm{b}$, the process identifier, and the local subset identifier:
\begin{align}
    \label{eq:param_dist_bw_b}
    \bm{b} &= w_b(\bm{p}, \bm{l})\\
    \label{eq:param_dist_fw_p}
    \bm{p} &= w_p(\bm{b})\\
    \label{eq:param_dist_l}
    \bm{l} &= w_l(\bm{b})
\end{align}
% Table~\ref{tab:symbols} summarizes the principal symbols and conventions used throughout the rest of this section.
% \alexnick{Since we only use block distributions, we can probably the process-local identifier from the above generic definition of a distribution. We may want to keep it though, just to show that we can "easily" extend our framework to use block-cyclic distributions if needed (for reviewers).}
\begin{table}[t]
 \footnotesize
 \centering
 \caption{Symbols and notations used in this section.}
  \rowcolors{2}{gray!15}{white}
 \begin{tabular}{p{.8cm}p{7cm}}
  \toprule
  \textbf{Name} & \textbf{Description} \\
\midrule
  $\bm{I}$       &   $N$-dimensional iteration space. Each dimension $j \in 0...N_j-1$ has size $I_j$\\
  $\bm{i}$  &   Vector of size $N$ representing the coordinates of an element of the iteration or data space \\
  $\bm{B}$  &   Vector of size $N$ representing the size of a partition of $\bm{I}$\\
  $\bm{b}$  &   Given a partition of $\bm{I}$, it is a vector of size $N$ representing the coordinates of the element that has the shortest Euclidean distance from the origin \\
  $\bm{o}$  &   Given an element of the space $\bm{i}$, it is a vector of size $N$ representing the offset coordinates of the element relative to $\bm{b}$\\
  $\bm{p}$  &   Unique process identifier \\
%  $\bm{P}$  &   ...<only used in block cyclic> \\
  $\bm{l}$  &   Process local partition identifier\\
  
  \midrule
  \hiderowcolors \multicolumn{2}{c}{\textbf{Mappings}}\\ \showrowcolors
  \midrule
  $u(\bm{i})$   & Returns the base element $\bm{b}$ of the partition $\bm{i}$ belongs to\\
  $v(\bm{i})$   & Returns the offset $\bm{o}$ of the partition $\bm{i}$ belongs to\\
  $w_b(\bm{p},\bm{l})$   & Returns the subset $\bm{b}$  that belongs to $\bm{p}$ and has the unique process-local identifier $\bm{l}$  \\
  $w_p(\bm{b})$   & Returns the process identifier to which $\bm{b}$ is assigned \\
  $w_l(\bm{b})$   & Returns the local subset identifier of $\bm{b}$ \\
%   \midrule
%   \hiderowcolors \multicolumn{2}{c}{\textbf{Conventions}}\\ \showrowcolors
%   \midrule
%   $\star^{(y)}$   &     Refers to the iteration space. It can be applied to all variables, vectors, and mappings previously defined \\
%   $\star^{(x)}$   &     Refers to the data space. It can be applied to all variables, vectors, and mappings previously defined \\ 
  \bottomrule
  \end{tabular}
  \label{tab:symbols}
  \vspace{-2em}
\end{table}

\subsection{Block Distribution}
\label{sec:block-distr}

We block-distribute the above space $\bm{I}$ in the following manner.
First, we select a constant block size, described by the vector $\bm{B} = (B_0, B_1, ..., B_{N-1})^T$, and we tile the space to $\smash[b]{\Pi_{j=0}^{N-1}\ceil*{I_j/B_j}}$ orthogonal blocks.
This results in a regular grid of size $\smash[b]{\ceil*{I_0/B_0} \times ... \times \ceil*{I_{N-1}/B_{N-1}}}$, where each block has coordinates $\bm{bi} = (bi_0, bi_1, ..., bi_{N-1})^T$, with $bi_j \in 0 ... \smash[b]{\ceil*{I_j/B_j} - 1}$.
Therefore, each block has the following unique identifier:
\begin{align}
    \label{eq:block_cyclic_b}
    \bm{b} = diag(\bm{B}) \cdot \bm{bi}
\end{align}
$diag(\bm{B})$ is the diagonal matrix, such that $diag(\bm{B})_{jj} = B_j$.
We then select the number of (MPI) processes $P$ and arrange them in a $N$-dimensional Cartesian grid, the size of which is described by the vector $\bm{P} = (P_0, P_1, ..., P_{N-1})^T$, with $P = \smash[b]{\Pi_{j=0}^{N-1}P_j}$.
Each process has grid coordinates $\bm{p} = (p_0, p_1, ..., p_{N-1})^T$, with $p_j \in 0 ... P_j - 1$.
% We distribute the blocks to the processes in a cyclic manner, so that each block with coordinates $(bi_0, bi_1, ..., bi_{D-1})^T$ is assigned to process $(bi_0 \textit{ mod } P_0, bi_1 \textit{ mod } P_1, ..., bi_{D-1} \textit{ mod } P_{D-1})^T$.
% The block-cyclic distribution for $1$-dimensional and $2$-dimensional spaces are shown in Fig.~\ref{fig:block-cyclic-1d} and \ref{fig:block-cyclic-2d} respectively.
% The cyclic assignment of the blocks results in a grid of $\smash[b]{\Pi_{j=0}^{D-1}\ceil*{S_j/\left(B_jP_j\right)}}$ \emph{courses}.
% Each course consists of $\smash[b]{\Pi_{j=0}^{D-1}P_j}$ blocks, one for each process, and has coordinates $\bm{l} = (l_0, l_1, ..., l_{D-1})^T$, with $l_j \in 0 ... \smash[b]{\ceil*{S_j/\left(B_jP_j\right)} - 1}$.
% This is equivalent to stating that each process is assigned $\smash[b]{\Pi_{j=0}^{D-1}\ceil*{S_j/\left(B_jP_j\right)}}$ blocks, one from each course, and defining $\bm{l}$ as the process-local block index vector.
% Based on the above, we rewrite the block coordinates as:
% \begin{align}
%     \label{eq:block_cyclic_bi}
%     \bm{bi} = diag(\bm{P}) \cdot \bm{l} + \bm{p}
% \end{align}
% $diag(\bm{P})$ is the diagonal matrix, such that $diag(\bm{P})_{jj} = P_j$.
We assign a single block to each process, so that each block with coordinates $(bi_0, bi_1, ..., bi_{N-1})^T$ is assigned to process $(p_0, p_1, ..., p_{N-1})^T$.
In other words, $\bm{bi} = \bm{p}$ and we rewrite the block identifier as:
\begin{align}
    \label{eq:block_cyclic_b_2}
    \bm{b} = diag(\bm{B}) \cdot \bm{p}
\end{align}

Substituting Eq.~(\ref{eq:block_cyclic_b_2}) on (\ref{eq:param_dist_idx}), we rewrite the $N$-dimensional index vector $\bm{i}$ as the affine expression:
\begin{equation}
\label{eq:block_cyclic_idx}
\bm{i} = diag(\bm{B}) \cdot \bm{p} + \bm{o}
\end{equation}
The index $\bm{i}$ is decomposed to (a) the grid-coordinates vector $\bm{p}$ of \revisionanotag{the process to which it is assigned}, and (b) the offset vector $\bm{o} = (o_0, o_1, ..., o_{N-1})^T$, with $o_j \in 0 ... B_j - 1$, that describes its coordinates relative to the beginning of \revisionanotag{the block to which it belongs.}
Eq.~\ref{eq:block_cyclic_idx} can be decomposed to $N$ independent affine expressions, one for each dimension:
\begin{align}
    \label{eq:block_cyclic_idx_per_dim}
    i_j = p_jB_j + o_j
\end{align}
% The coordinates $l_j$, $p_j$, and $o_j$ can be computed based on $i_j$:
The mappings of Eqs.~(\ref{eq:param_dist_b}),~(\ref{eq:param_dist_o}),~(\ref{eq:param_dist_fw_p}) for the block distribution are given per dimension as follows:
\begin{align}
    \label{eq:b_coord}
    b_j &= u(i_j) = B_j\floor*{\frac{i_j}{B_j}}\\
    \label{eq:o_coord}
    o_j &= v(i_j) = i_j \textit{ mod } B_j\\
    \label{eq:p_coord}
    p_j &= w_p(b) = \frac{b_j}{B_j} = \floor*{\frac{i_j}{B_j}}
\end{align}
% Nested integer divisions are merged as shown in Eq.~\ref{eq:l_coord}, according to Theorem 3.10 from \cite{graham1994concrete}, with $f(x) = x/P_j$ and $x = i_j/B_j$.
We note that, in the block distribution, the mappings related to the local subsets are irrelevant, since each process is assigned a single block and the local subset identifier is the zero vector.

% If we describe a unified space $I = \smash[b]{\bigtimes_{j=0}^{D-1}\{0...S_j-1\}}$ as a tuple $\left(S_0, S_1, ..., S_{D-1}\right)$ of the dimension lengths, then its block distribution can be described with \textit{reshaping}.
% The total size of the space is immutable, and therefore, this reshaping can be implicit.
% We choose to make it explicit to facilitate the analysis of the data movement:
% \begin{align}
%     \left(S_0, S_1, ..., S_{D-1}\right) \rightarrow \left(P_0, ..., P_{D-1}, B_0, ..., B_{D-1}\right) \nonumber
% \end{align}
% Similarly, the space coordinates are rewritten as tuples of the process identifier $\bm{p}$, and the offset coordinates $\bm{o}$:
% \begin{align}
%     \bm{i} = \left(i_0, i_1, ..., i_{D-1}\right)^T \rightarrow \left(p_0, ..., p_{D-1}, o_0, ..., o_{D-1}\right)^T \nonumber
% \end{align}

\begin{table*}[h!]
 \footnotesize
 \centering
 \caption{List of benchmarks executed, together with their algebraic, and Einstein summation notations.}
  \label{tab:benchmarks}
  \rowcolors{2}{gray!15}{white}
 \begin{tabular}{p{2.1cm}p{5cm}p{5.3cm}p{4.05cm}}
  \toprule
  \textbf{Name} & \textbf{Algebraic Notation} & \textbf{Definitions} & \textbf{Einstein Summation} \\
  \midrule
  \midrule
  \hiderowcolors \multicolumn{4}{c}{\textbf{Matrix-Matrix products}}\\ \showrowcolors
  \midrule
  \texttt{1MM}          &   $\bm{A} \cdot \bm{B}$
                        &   $\bm{A} \in \mathbb{C}^{I_0 \times I_1}, \bm{B} \in \mathbb{C}^{I_1 \times I_2}$
                        &   \texttt{ij,jk->ik}\\
  \texttt{2MM}          &   $\bm{A} \cdot \bm{B} \cdot \bm{C}$
                        &   $\bm{C} \in \mathbb{C}^{I_2 \times I_3}$
                        &   \texttt{ij,jk,kl->il}\\
  \texttt{3MM}          &   $\bm{A} \cdot \bm{B} \cdot \bm{C} \cdot \bm{D}$
                        &   $\bm{D} \in \mathbb{C}^{I_3 \times I_4}$
                        &   \texttt{ij,jk,kl,lm->im}\\
  \midrule
  \hiderowcolors \multicolumn{4}{c}{\textbf{Matricized Tensor times Khatri-Rao products}}\\ \showrowcolors
  \midrule
  \texttt{MTTKRP-O3-M0} &   $\bm{\mathcal{X}} \times_{0} \left(\bm{U}^{1}\odot\bm{U}^{2}\right)$
                        &   $\bm{\mathcal{X}} \in \mathbb{C}^{I_0 \times I_1 \times I_2}, \bm{U}^{n} \in \mathbb{C}^{I_n \times R}$
                        &   \texttt{ijk,ja,ka->ia}\\
  \texttt{MTTKRP-O3-M1} &   $\bm{\mathcal{X}} \times_{1} \left(\bm{U}^{0}\odot\bm{U}^{2}\right)$
                        &   
                        &   \texttt{ijk,ia,ka->ja}\\
  \texttt{MTTKRP-O3-M2} &   $\bm{\mathcal{X}} \times_{2} \left(\bm{U}^{0}\odot\bm{U}^{1}\right)$
                        &   
                        &   \texttt{ijk,ia,ja->ka}\\
  \texttt{MTTKRP-O5-M0} &   $\bm{\mathcal{X}} \times_{0} \left(\bm{U}^{1}\odot\bm{U}^{2}\odot\bm{U}^{3}\odot\bm{U}^{4}\right)$
                        &   $\bm{\mathcal{X}} \in \mathbb{C}^{I_0 \times I_1 \times I_2 \times I_3 \times I_4}, \bm{U}^{n} \in \mathbb{C}^{I_n \times R}$
                        &   \texttt{ijklm,ja,ka,la,ma->ia}\\
  \texttt{MTTKRP-O5-M2} &   $\bm{\mathcal{X}} \times_{2} \left(\bm{U}^{0}\odot\bm{U}^{1}\odot\bm{U}^{3}\odot\bm{U}^{4}\right)$
                        &   
                        &   \texttt{ijklm,ia,ja,la,ma->ka}\\
  \texttt{MTTKRP-O5-M4} &   $\bm{\mathcal{X}} \times_{4} \left(\bm{U}^{0}\odot\bm{U}^{1}\odot\bm{U}^{2}\odot\bm{U}^{3}\right)$
                        &   
                        &   \texttt{ijklm,ia,ja,ka,la->ma}\\
  \midrule
  \hiderowcolors \multicolumn{4}{c}{\textbf{Tensor times Matrix Chain}}\\ \showrowcolors
  \midrule
%   \texttt{TTMc-O3-M0}   &   $\bm{\mathcal{X}} \times_{1} \bm{U}^{1} \times_{1} \bm{U}^{2}$
%                         &   $\bm{\mathcal{X}} \in \mathbb{C}^{I_0 \times I_1 \times I_2}, \bm{U}^{n} \in \mathbb{C}^{I_n \times R_n}$
%                         &   \texttt{ijk,jb,kc->ibc}\\
  \rowcolor{gray!15}
  \texttt{TTMc-O5-M0}   &   $\bm{\mathcal{X}} \times_{1} \bm{U}^{1} \times_{2} \bm{U}^{2} \times_{3} \bm{U}^{3} \times_{4} \bm{U}^{4}$
                        &   $\bm{\mathcal{X}} \in \mathbb{C}^{I_0 \times I_1 \times I_2 \times I_3 \times I_4}, \bm{U}^{n} \in \mathbb{C}^{I_n \times R_n}$
                        &   \texttt{ijklm,jb,kc,ld,me->ibcde}\\
  \bottomrule
  \end{tabular}
  \vspace{-1.5em}
\end{table*}

\subsection{Redistributing Data}
\label{sec:redistr}

Since a \revisionanotag{multilinear algebra kernel} may be decomposed into groups of statements as described in Sec.~\ref{sec:multi-bound} and these statements may be distributed with different block sizes, redistribution of data may be needed.
Let there be two groups of statements with iteration (sub-)spaces $\bm{I}^{(x)}$ and $\bm{I}^{(y)}$ and an $N$-mode tensor $\bm{\mathcal{X}}$ that resides on both of them.
These spaces have dimensionality $N^{(x)}, N^{(y)} \geq N$ and \revisionanotag{their intersection is a superset} of the exact vector space of $\bm{\mathcal{X}}$.
% which implies data replication. \alexnick{Data replication must also be described in more detail somewhere.}
Without loss of generality, for the purposes of the following data movement analysis, we consider only the subsets of those spaces that coincide with the vector space of $\bm{\mathcal{X}}$.
If the spaces have identical distributions, i.e., they are characterized by the same Cartesian process grids and block sizes, then no redistribution is needed.
However, if the distributions are not the same, then $\bm{\mathcal{X}}$ needs to be redistributed.

Copying the data from one distribution to the other is straightforward in a \textit{per-element} manner.
Using Eq.~(\ref{eq:block_cyclic_idx}), we decompose the index coordinate of each tensor element to the block sizes, process identifier, and offset coordinates that correspond to each distribution:
\begin{align}
    \label{eq:block_cyclic_id}
    \bm{i} &= diag(\bm{B}^{(x)}) \cdot \bm{p}^{(x)} + \bm{o}^{(x)}\nonumber\\
           &= diag(\bm{B}^{(y)}) \cdot \bm{p}^{(y)} + \bm{o}^{(y)}
\end{align}
The per-dimension process and offset coordinates are computed using Eqs.~(\ref{eq:o_coord}),~(\ref{eq:p_coord}).
This information makes it possible to establish one-side communication and copy the data from one distribution to the other, one element at a time.

Naturally, message aggregation is a vital optimization step to reduce communication overheads by coalescing individual communication requests in fewer but larger messages. 
We analyze the data movement needed for redistributing a single subset of data $\bm{b}^{(x)}$ to the $y$-distribution.
We partition the block to $k$ (disjoint) partitions so that for each partition, communication is needed only with some (other) partition of a single subset $\bm{b}^{(y)}$ from the second distribution.
To find those partitions, we rewrite the index coordinates using Eq.~(\ref{eq:param_dist_idx}):
\begin{align}
    \label{eq:param_dist_one_sided_aggr_b}
    \bm{i}^{(x)} = \bm{i}^{(y)} = \bm{b}^{(y)} + \bm{o}^{(y)}
\end{align}
% To find those partitions, we combine Eq.~(\ref{eq:block_cyclic_id}) with Eq.~(\ref{eq:b_coord}) and we derive a per-dimension mapping between the $y$-distribution's offset coordinates $\bm{o}^{(y)}$ and the $x$-distribution's block identifier $\bm{b}^{(x)}$:
% \begin{align}
%     p_j^{(x)}B_j^{(x)} &= u^{(x)}\left(p_j^{(y)}B_j^{(y)} +o_j^{(y)}\right)\nonumber\\
%     \label{eq:param_dist_one_sided_aggr_b}
%     &= B_j^{(x)}\floor*{\frac{p_j^{(y)}B_j^{(y)} +o_j^{(y)}}{B_j^{(x)}}}
% \end{align}
We can consider $\bm{b}^{(x)}$ to be a step function of $\bm{o}^{(y)}$.
% Therefore, depending on the form of the data access method $f$, we can analyze Eq.~(\ref{eq:param_dist_one_sided_aggr_b}) and find symbolic expressions for the values of $\bm{b}^{(x)}$.
Therefore, the solution has the form:
\begin{align}
    \label{eq:param_dist_one_sided_range_b}
    \bm{b}^{(x)} =
    \begin{cases}
        \bm{b}_0, &o^{(y)} \in \textit{partition}_0^{(y)}\\
        \bm{b}_1, &o^{(y)} \in \textit{partition}_1^{(y)}\\
        ...\\
        \bm{b}_{k-1}, &o^{(y)} \in \textit{partition}_{k-1}^{(y)}\\
    \end{cases}
\end{align}
Similarly, combining Eq.~(\ref{eq:param_dist_o}) with Eq.~(\ref{eq:param_dist_b}), we construct a mapping between the $y$-distribution offset coordinates $\bm{o}^{(y)}$ and the $x$-distribution offset coordinates $\bm{o}^{(x)}$: 
\begin{align}
    \label{eq:param_dist_one_sided_aggr_o}
    \bm{o}^{(x)} = \left(v^{(x)} \circ f\right)\left(\bm{b}^{(y)} + \bm{o}^{(y)}\right)
\end{align}
Using the partitions of $\bm{o}^{(y)}$ found in Eq.~(\ref{eq:param_dist_one_sided_range_b}), we find the corresponding partitions of $\bm{o}^{(x)}$:
\begin{align}
    \label{eq:param_dist_one_sided_range_o}
    \bm{o}^{(x)} \in
    \begin{cases}
        \textit{partition}_0^{(x)}, &o^{(y)} \in \textit{partition}_0^{(y)}\\
        \textit{partition}_1^{(x)}, &o^{(y)} \in \textit{partition}_1^{(y)}\\
        ...\\
        \textit{partition}_{k-1}^{(x)}, &o^{(y)} \in \textit{partition}_{k-1}^{(y)}\\
    \end{cases}
\end{align}
In general, we expect the number of partitions $k$ to be a function of the subset sizes $\bm{B}^{y}$, $\bm{B}^{x}$.
%, and the access method $f$.

We construct Eqs.~(\ref{eq:param_dist_one_sided_aggr_b}) and (\ref{eq:param_dist_one_sided_aggr_o}) for the block distribution per dimension (the dimension subscript $j$ is omitted for brevity):
\begin{align}
    \label{eq:block_cyclic_one_sided_aggr_b}
    p^{(x)} &= \floor*{\frac{p^{(y)}B^{(y)} + o^{(y)}}{B^{(x)}}}\\
    \label{eq:block_cyclic_one_sided_aggr_o}
    o^{(x)} &= p^{(y)}B^{(y)} + o^{(y)} \textit{ mod } B^{(x)}
\end{align}
Using the $x \textit{ mod } y = x - y\floor*{x/y}$ property of the modulo operation, for $x$ integer and $y$ positive integer, we rewrite Eq.~(\ref{eq:block_cyclic_one_sided_aggr_o}):
\begin{align}
    \label{eq:block_cyclic_one_sided_aggr_o2}
    o^{(x)} &= p^{(y)}B^{(y)} + o^{(y)} - B^{(x)}\floor*{\frac{p^{(y)}B^{(y)} + o^{(y)}}{B^{(x)}}}
\end{align}
% Analyzing the data movement depends on the form of the data access method $f$.
% We provide a general solution for $f(i) = \smash[b]{\floor*{(\alpha i + \beta)/\gamma}}$, with $\alpha$, $\gamma$ positive integers, and $\beta$ integer.
% Such an access method covers most of the linear algebra kernels, solvers, and integer division, frequently found in many algorithms:
% Therefore, we rewrite Eqs.~(\ref{eq:block_cyclic_one_sided_aggr_b}) and (\ref{eq:block_cyclic_one_sided_aggr_o2}) as:
% \begin{align}
%     \label{eq:block_cyclic_one_sided_aggr_b_replf}
%     &l^{(x)}P^{(x)} + p^{(x)} = \floor*{\frac{\alpha \left(l^{(y)}P^{(y)}+p^{(y)}\right)B^{(y)} + \alpha o^{(y)} + \beta}{\gamma B^{(x)}}}\\
%     \label{eq:block_cyclic_one_sided_aggr_o_replf}
%     &o^{(x)} = \floor*{\frac{\alpha \left(l^{(y)}P^{(y)}+p^{(y)}\right)B^{(y)} + \alpha o^{(y)} + \beta}{\gamma}} - B^{(x)}\floor*{\frac{\alpha \left(l^{(y)}P^{(y)}+p^{(y)}\right)B^{(y)} + \alpha o^{(y)} + \beta}{\gamma B^{(x)}}}
% \end{align}
\revisionanotag{The floor division $\smash[b]{\floor*{\left( p^{(y)}B^{(y)} + o^{(y)}\right)/B^{(x)}}}$ appears on both Eqs.~(\ref{eq:block_cyclic_one_sided_aggr_b}),~(\ref{eq:block_cyclic_one_sided_aggr_o2}).
To facilitate the study of the values that this expression takes, we rewrite the block identifier $p^{(y)}B^{(y)}$ in terms of the denominator, introducing auxiliary non-negative integer variables $\xi$ and $\lambda$. $\xi$ is the quotient of the division between $p^{(y)}B^{(y)}$ and $B^{(x)}$, while $\lambda$ is the remainder:}
% The expression $\smash[b]{\floor*{\left( p^{(y)}B^{(y)} + o^{(y)}\right)/B^{(x)}}}$ appears on both Eqs.~(\ref{eq:block_cyclic_one_sided_aggr_b}),~(\ref{eq:block_cyclic_one_sided_aggr_o2}).
% To study the values this expression takes for different values of $o^{(y)}$, we rewrite part of the numerator as:
\begin{align}
    \label{eq:xi_lambda}
    p^{(y)}B^{(y)} &= \xi B^{(x)} + \lambda\\
% \end{align}
% % with:
% \begin{align}
    \label{eq:xi}
    \xi &= \floor*{\frac{p^{(y)}B^{(y)}}{B^{(x)}}} \in \mathbb{N}\\
    \label{eq:lambda}
    \lambda &= p^{(y)}B^{(y)} \textit{ mod } B^{(x)} \in 0 ... B^{(x)} - 1
\end{align}
Using Eq.~(\ref{eq:xi_lambda}), we rewrite Eq.~(\ref{eq:block_cyclic_one_sided_aggr_b}) as:
\begin{align}
    \label{eq:block_cyclic_one_sided_aggr_b_ranges}
    p^{(x)} &= \floor*{\frac{\xi B^{(x)} + \lambda + o^{(y)}}{B^{(x)}}} = \xi + \floor*{\frac{\lambda + o^{(y)}}{B^{(x)}}} \nonumber\\
    &= \xi +
    \begin{cases}
        0, & 0 \leq o^{(y)} < B^{(x)} - \lambda\\
        1, & B^{(x)} - \lambda \leq o^{(y)} < 2 B^{(x)} - \lambda\\
        ...\\
        k-1, & (k - 1) B^{(x)} - \lambda \leq o^{(y)} < B^{(y)}\\
    \end{cases}
\end{align}
where:
\begin{align}
    \label{eq:kappa}
    (k - 1)B^{(x)} - \lambda < B^{(y)} \Leftrightarrow k - 1 < \ceil*{\frac{B^{(y)} + \lambda}{B^{(x)}}} \nonumber\\
    \Rightarrow k \equiv \ceil*{\frac{B^{(y)} + \lambda}{B^{(x)}}} \leq \ceil*{\frac{B^{(y)} + B^{(x)} - 1}{B^{(x)}}} \nonumber\\
    \Rightarrow k \leq \ceil*{\frac{B^{(y)} - 1}{B^{(x)}}} + 1
\end{align}
Substituting Eqs.~(\ref{eq:xi_lambda}) and (\ref{eq:block_cyclic_one_sided_aggr_b_ranges}) in Eq.~(\ref{eq:block_cyclic_one_sided_aggr_o2}), we find the corresponding ranges for $o^{(x)}$:
\begin{align}
    o^{(x)} &= \xi B^{(x)} + \lambda + o^{(y)} - B^{(x)}\left(\xi + \floor*{\frac{\lambda + o^{(y)}}{B^{(x)}}}\right) \nonumber\\
    &=\lambda + o^{(y)}  - B^{(x)}\floor*{\frac{\lambda + o^{(y)}}{B^{(x)}}} \nonumber\\
    \label{eq:block_cyclic_one_sided_aggr_o_ranges}
    &\in
    \begin{cases}
        \left[\lambda,  B^{(x)}\right), \quad 0 \leq o^{(y)} < B^{(x)} - \lambda\\
        \left[0,  B^{(x)}\right), \quad B^{(x)} - \lambda \leq o^{(y)} < 2B^{(x)} - \lambda\\
        ...\\
        \left[0,  \lambda + B^{(y)} - (k-1)B^{(x)}\right), \\
        \qquad(k-1)B^{(x)} - \lambda \leq o^{(y)} < B^{(y)}
    \end{cases}
\end{align}
% The exact values, that $o^{(x)}$ takes in the above ranges, depend on the values of $\alpha$ and $\gamma$.
% If $\gamma = 1$, then the ranges have a fixed stride equal to $\alpha$, otherwise the stride is not constant.

% We note that Eq.~\ref{eq:block_cyclic_one_sided_aggr_b} can be used to solve message matching by substituting $o^{(y)}$ with its maximum ($B^{(y)}$) and minimum values ($0$):
We note that Eq.~\ref{eq:block_cyclic_one_sided_aggr_b} can be used to solve message matching by substituting $o^{(y)}$ with its minimum and maximum values:
\begin{align}
    \label{eq:msg-match}
    % p^{(x)} &= \floor*{\frac{p^{(y)}B^{(y)} + o^{(y)}}{B^{(x)}}}\nonumber\\
    % \ceil*{\frac{p^{(x)}B^{(x)} - o^{(y)}}{B^{(y)}}} \leq p^{(y)} &< \ceil*{\frac{(p^{(x)} + 1)B^{(x)} - o^{(y)}}{B^{(y)}}}\nonumber\\
    \ceil*{\frac{p^{(x)}B^{(x)} + 1}{B^{(y)}}} - 1\leq p^{(y)} &< \ceil*{\frac{(p^{(x)} + 1)B^{(x)}}{B^{(y)}}}
\end{align}
% In Eq.~\ref{eq:msg-match} we substitute $o^{(y)}$ with the maximum ($B^{(y)}$) and minimum value ($0$).
Using this formula, each $x$-distribution process performs a loop over candidate $y$-distribution processes to which it may need to send data, allowing the implementation of redistribution with two-sided communication.

\section{Evaluation}
\label{sec:evaluation}

\begin{figure*}[h]
    \centering
    \includegraphics[width=.95\textwidth]{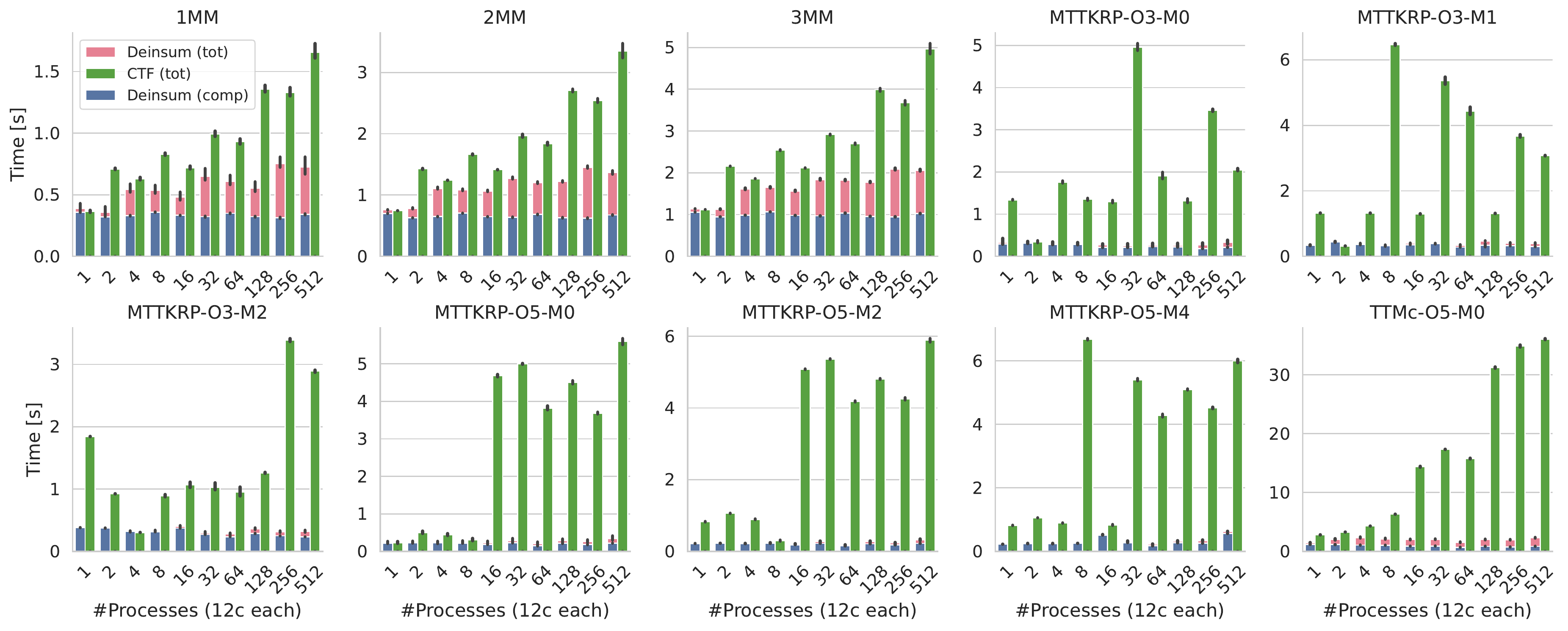}
    \caption{Deinsum and CTF CPU runtimes on up to 512 nodes. Deinsum's computation time is also shown as part of the total runtime.}
    \label{fig:results-placeholder}
    \vspace{-1.5em}
\end{figure*}

We evaluate the performance of the codes generated by our framework using the benchmarks described in Tab.~\ref{tab:benchmarks}.
We start with matrix multiplications; a single product (\texttt{1MM}), a chain of two (\texttt{2MM}), and three products (\texttt{3MM}).
We proceed with higher-order tensor operations, specifically MTTKRP with order-3 and -5 tensors; and order-5 TTMc.
We perform weak scaling experiments using the initial problem sizes (for single-node execution) and scaling factors presented in Tab.~\ref{tab:weak_scaling}.

% We compare the runtime and parallel efficiency of our framework against the Cyclops Tensor Framework (CTF).
% We compute the parallel efficiency of our framework using as baseline the execution on a single-node of an implementation where all MPI- and data redistribution-related calls are removed.
% Therefore, the parallel efficiency on a single node may not necessarily be 100\% due to false communication overhead.
% However, we note that, in some cases, it is possible to achieve higher than 100\% parallel efficiency in larger node counts.
% In general, this occurs due to the distribution of the data leading to sub-computational kernels with higher data reuse compared to the execution on a single node.

\begin{table}[h!]
 \footnotesize
 \caption{List of benchmarks, initial problem sizes, and scaling factors as a function of the number of processes $P$.}
 \centering
%   \rowcolors{2}{gray!15}{white}
 \begin{tabular}{p{2.8cm}p{3.5cm}p{1cm}}
  \toprule
  \textbf{Benchmark} & \textbf{Initial Problem Size} & \textbf{Scaling} \\
  \midrule
  % We don't need those, they are given in Table IV already
  %\midrule
  %\hiderowcolors \multicolumn{3}{c}{\textbf{Matrix-Matrix products}}\\ \showrowcolors
  %\midrule
  \rowcolor{gray!15}   \texttt{1MM}          &   $I^n = 4096, n \in 0..2$ &   $\sqrt[3]{P}$\\
                      \texttt{2MM}          &   $I^n = 4096, n \in 0..3$ &   $\sqrt[3]{P}$\\
  \rowcolor{gray!15}   \texttt{3MM}          &   $I^n = 4096, n \in 0..4$ &   $\sqrt[3]{P}$\\                      
  %\midrule
  %\hiderowcolors \multicolumn{3}{c}{\textbf{Matricized Tensor times Khatri-Rao products}}\\ \showrowcolors
  %\midrule
  
                        &   $I^n = 1024, n \in 0..2$ &  $\sqrt[4]{P}$\\
    \multirow{-2}{*}{\texttt{MTTKRP-03-M\{0,1,2\}}}   &   $R = 24$ &   $\sqrt[4]{P}$ \\
    \rowcolor{gray!15}
                        &   $I^n = 1024, n \in 0..4$ &  $\sqrt[6]{P}$\\
    \rowcolor{gray!15} \multirow{-2}{*}{\texttt{MTTKRP-05-M\{0,2,4\}}}   &   $R = 24$ &   $\sqrt[6]{P}$\\
  %\midrule
  %\hiderowcolors \multicolumn{3}{c}{\textbf{Tensor times Matrix Chain}}\\ \showrowcolors
  %\midrule
                      &   $I^n = 60, n \in 0..4$  &   $\sqrt[6]{P}$\\
  \multirow{-2}{*}{\texttt{TTMc-05-M0}}   &   $R^n = 24, n \in 0..4$  &   $\sqrt[6]{P}$\\
  \bottomrule
  \end{tabular}
  \vspace{-1.5em}
  \label{tab:weak_scaling}
\end{table}

\subsection{Experimental Setup}

We run the benchmarks on the Piz Daint supercomputer, up to 512 nodes.
\revisionanotag{Each Cray XC50 compute node has a 12-core Intel E5-2690 v3 CPU @ 2.6Ghz, an Nvidia P100 GPU with 16GB of memory, and 64GB of main memory.}
The nodes are connected through a Cray Aries network using a Dragonfly topology.
\revisionanotag{For CPU execution}, we test the latest verified version of CTF (commit ID c4f89dc~\cite{ctf-repo}) from its GitHub repository.
The CTF C++ codes and those auto-generated by Deinsum are compiled with GCC version 9.3.0 and linked against the same libraries; Cray MPICH \revisionanotag{CUDA-aware} 7.7.18 for MPI communication, and Intel oneAPI MKL 2021.3.0 for BLAS support.
Furthermore, both CTF and Deinsum utilize the High-Performance Tensor Transpose library (HPTT)~\cite{hptt} for out-of-place tensor transpositions.
Deinsum codes are linked against the latest version of HPTT (commit ID 9425386~\cite{hptt-repo}) from its GitHub repository.
CTF automatically downloads and compiles a forked version of HPTT (commit ID 3c77169~\cite{hptt-edgar-repo}).
\revisionanotag{For GPU execution, we test CTF's \texttt{gpu\_devel\_v2} branch (commit ID 0c41739b).
Deinsum utilizes cuTENSOR~\cite{cutensor} for single-GPU binary tensor operations. All GPU programs are compiled using NVCC and CUDA 11.0.}

% \begin{figure*}[h]
%     \centering
%     \includegraphics[width=.95\textwidth]{fig2.pdf}
%     \caption{Deinsum and CTF CPU runtimes on up to 512 nodes. Deinsum's computation time is also shown as part of the total runtime.}
%     \label{fig:results-placeholder}
%     \vspace{-1.5em}
% \end{figure*}

\subsection{CPU Results}

\revisionanotag{We compare Deinsum's performance with CTF's on CPU.}
For each benchmark and framework, we measure the runtime of at least ten executions, and we plot the median and the 95\% confidence interval using bootstrapping~\cite{efronbootstrap}.
The results are shown in Fig~\ref{fig:results-placeholder}.
The blue and pink bars together indicate Deinsum's runtime.
The blue bar corresponds to the compute runtime, including any necessary intra-node tensor transpositions, which we measure by running a version of the code stripped of any inter-node communication for each benchmark.
The pink bar represents the communication overhead, which we estimate by subtracting the compute runtime from the total execution.
The green bar shows CTF's execution.

% \begin{figure*}[h]
%     \centering
%     \includegraphics[width=.95\textwidth]{fig2.pdf}
%     \caption{Deinsum and CTF CPU runtimes on up to 512 nodes. Deinsum's computation time is also shown as part of the total runtime.}
%     \label{fig:results-placeholder}
%     \vspace{-1em}
% \end{figure*}

% \begin{figure}[h]
%     \centering
%     \includegraphics[width=.95\columnwidth]{fig3.pdf}
%     \caption{DaCe with and without SOAP on TTMc\_O5\_M0.}
%     \label{fig:results-placeholder}
% \end{figure}

All three matrix-matrix products exhibit similar scaling behavior.
The compute time is flat since it depends purely on the performance of the BLAS (MKL) GEMM kernel on the machine.
Deinsum's communication overhead increases in steps at 4, 32, and 256 nodes, especially in \texttt{1MM}.
This results from the SOAP-generated distribution on a three-dimensional process grid $(P_0, P_1, P_2)$.
The output product is partitioned into $P_0P_2$ blocks, with each block further split into $P_1$ partial sums.
In all node counts where the communication overhead increases, $P_1$ doubles.
For example, the process grid generated for 16 nodes is $(2, 2, 4)$ and for 32 nodes the size is $(2, 4, 4)$.
Therefore, the number of output blocks remains the same, while the block size increases due to weak scaling, and the depth of each \texttt{MPI\_Allreduce} doubles, potentially further increasing the latency of the operation.
CTF also exhibits a runtime increase in steps but at different node counts, implying a different distribution scheme.
On 512 nodes, Deinsum's speedups over CTF are $2.42\times$, $2.45\times$, and $2.38\times$ for each of the three (1MM,2MM,3MM) kernels.
Deinsum scales exceptionally well on the MTTKRP benchmarks exhibiting low communication overhead.
The speedups against CTF on 512 nodes range from $6.75$ to $19.00\times$.
The performance improvements on TTMc are $15.95\times$ on 512 nodes.
% \alexnick{Shall we push here a discussion on the performance of redistribution on high-order tensors or no need?}

% \begin{figure*}[t]
%     \centering
%     \includegraphics[width=.95\textwidth]{gpu.pdf}
%     \caption{\revisionanotag{Deinsum and CTF GPU runtimes on up to 512 nodes. Deinsum's runtime with input data resident in global GPU memory is also shown as part of the total runtime.}}
%     \vspace{-1.5em}
%     \label{fig:results-gpu}
% \end{figure*}

\subsection{GPU Results}

\begin{figure*}[t]
    \centering
    \includegraphics[width=.95\textwidth]{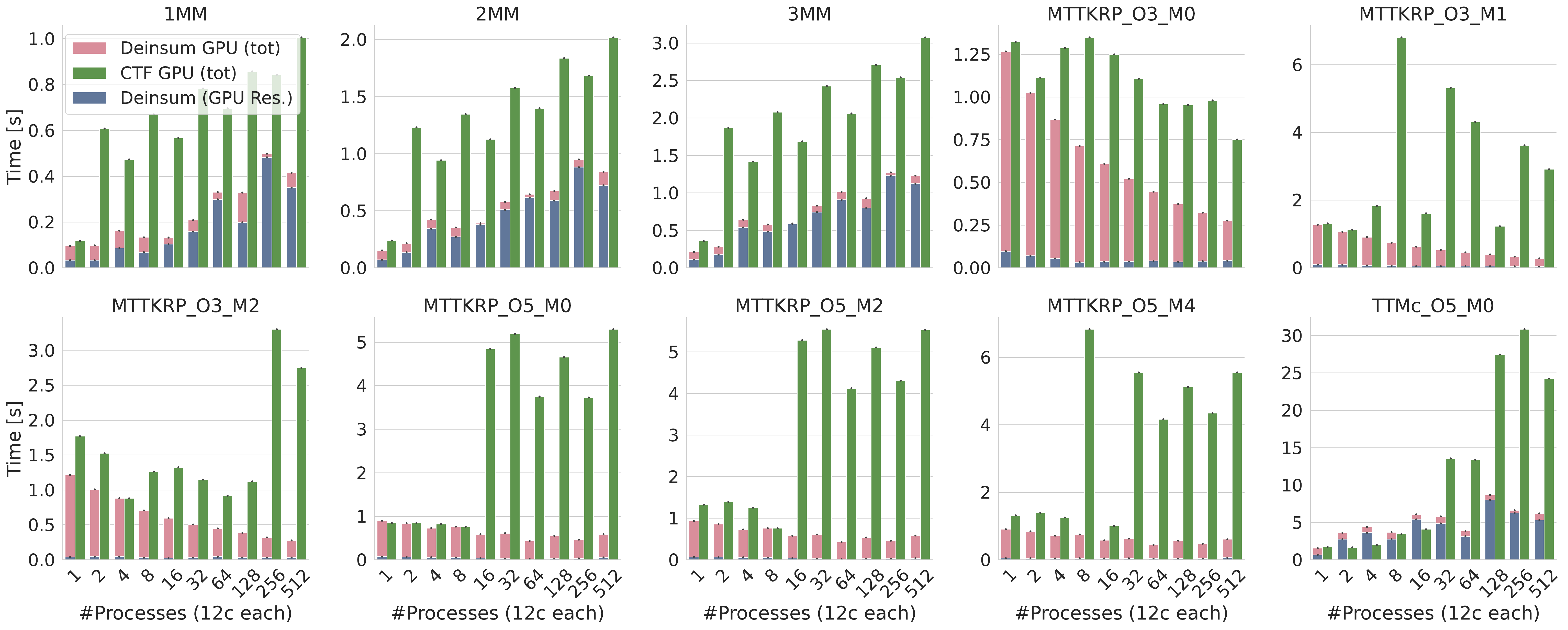}
    \caption{\revisionanotag{Deinsum and CTF GPU runtimes on up to 512 nodes. Deinsum's runtime with input data resident in global GPU memory is also shown as part of the total runtime.}}
    \vspace{-1.5em}
    \label{fig:results-gpu}
\end{figure*}

\revisionanotag{We compare Deinsum's performance with CTF's on GPU, using the same statistical methods as for CPU.
The results are shown in Fig.~\ref{fig:results-gpu}.
We make a distinction here among executions that utilize the GPU as an accelerator, i.e, the input and output data must be copied from/to the host to/from the device, and executions where the required data are already resident in global GPU memory and the output does not need to be copied back to main memory.
This distinction allows us to make an apples-to-apples comparison against CTF, which supports only the first execution type, while also showcasing Deinsum's performance on the second execution type, which may be more common in large applications ran on modern GPUs with dozens of GBs of memory.
The blue and pink bars together indicate Deinsum's runtime.
The blue bar is Deinsum GPU-resident execution, while the pink bar is the time required to copy the input and output data between the host and the device.
The green bar shows CTF's execution.
Overall, we are seeing similar performance trends as in the CPU execution.}

\section{Related Work}

In this section, we summarize prior work related to our main contributions.

\subsection{Multilinear Algebra Frameworks}

To the best of our knowledge, the only other framework that supports automated distribution and execution of arbitrary einsums in distributed memory machines is the Cyclops Tensor Framework (CTF)~\cite{ctf}.
TiledArray~\cite{tiledarray} is also a distributed framework that facilitates the composition of high-performance tensor arithmetic, but the user must explicitly program the data distribution into processes.
There exist many frameworks that execute arbitrary einsums in shared memory:
Apart from NumPy and the Optimized Einsum Python module, there are the Tensor Contraction Library (TCL) and Code Generator (TCCG)~\cite{gett}, and TBLIS~\cite{tblis} libraries that execute tensor operations and contractions on CPU.
The latter led to the development of cuTENSOR~\cite{cutensor}, an Nvidia GPU-compatible library for tensor contraction, reduction, and elementwise operations. \revisionanotag{cuTENSOR also supports multi-GPU setups utilizing NVLink via the cuTENSORMg API.}

\subsection{I/O Complexity Analysis}
Rigorous I/O complexity analysis dates back to the seminal work by Hong and Kung~\cite{redblue} who derived the first asymptotic I/O lower bound for a series of algorithms - among others, a classical matrix multiplication kernel. 
Their red-blue pebble game, underpinned by a two-level memory model, was extended multiple times to cover block accesses, kernel composition, and multiple memory levels~\cite{externalMem, redbluewhite,redredblue}. The data movement model used in this paper is due to Kwasniewski et al.~\cite{soap}, which is also based on the red-blue pebble game.
Other works that focus on the I/O complexity of linear algebra use variants of the discrete Loomis-Whitney inequality~\cite{loomisWhitney, irony},  Holder-Brascamp-Lieb inequalities~\cite{hbl1}, or recursion-based arguments~\cite{25d} to bound the I/O cost and derive communication avoiding schedules for series of linear algebra kernels.
There is significantly less work on the I/O complexity of multilinear algebra kernels.
Ballard et al. established a first parallel I/O lower bound for the order-$n$ MTTKRP~\cite{ballard2018communication}.
However,  their model prohibits decomposing the kernel into a series of binary contractions.

\subsection{Automated Data Distributions and Redistribution}
Automated (re)distribution algorithms similar to the analysis presented in Sec.~\ref{sec:distribution} are also employed by CTF.
Petitet et al.~\cite{petitet} have presented algorithmic redistribution methods for block-cyclic distributions.
Furthermore, considerable work in automating and optimizing the communication needed for multilinear algebra has been done by High Performance Fortran (HPF) compilers.
We categorize it into three different approaches: (a) via linear algebraic methods to construct symbolic expressions~\cite{ibm, ecoledesmines}; (b) using compile-time or runtime generated tables to store critical information, such as array access strides or communication mappings~\cite{rice1, rice2, rice3, riacs, lsu}; or (c) using the array slice expressions as index sets, such that the local and communication sets are described in terms of set operations, for example, unions and intersections~\cite{cmu, osu}.

\section{Conclusion}
We present Deinsum, a framework for automatic and near I/O optimal distribution of \revisionanotag{multilinear algebra kernels} expressed in Einstein notation.
Deinsum leverages the strength of the SOAP theoretical framework to derive a $6\times$ improved lower bound for MTTKRP, the main computational bottleneck of the CP decomposition.
Moreover, Deinsum vastly improves on CTF, the current state-of-the-art tensor computation framework, by up to $19\times$ on 512 nodes; the geometric mean of all observed speedups is $4.18\times$.
These results further solidify the validity of the improved tight I/O lower bounds described in Sec.~\ref{sec:single-bound} and confirm that the SOAP analysis provides not only theoretical but also tangible improvements to distributed computations.

\section{Acknowledgments}

This work received EuroHPC-JU funding with support from the European Union’s Horizon 2020 program and from the European Research Council \includegraphics[height=1em]{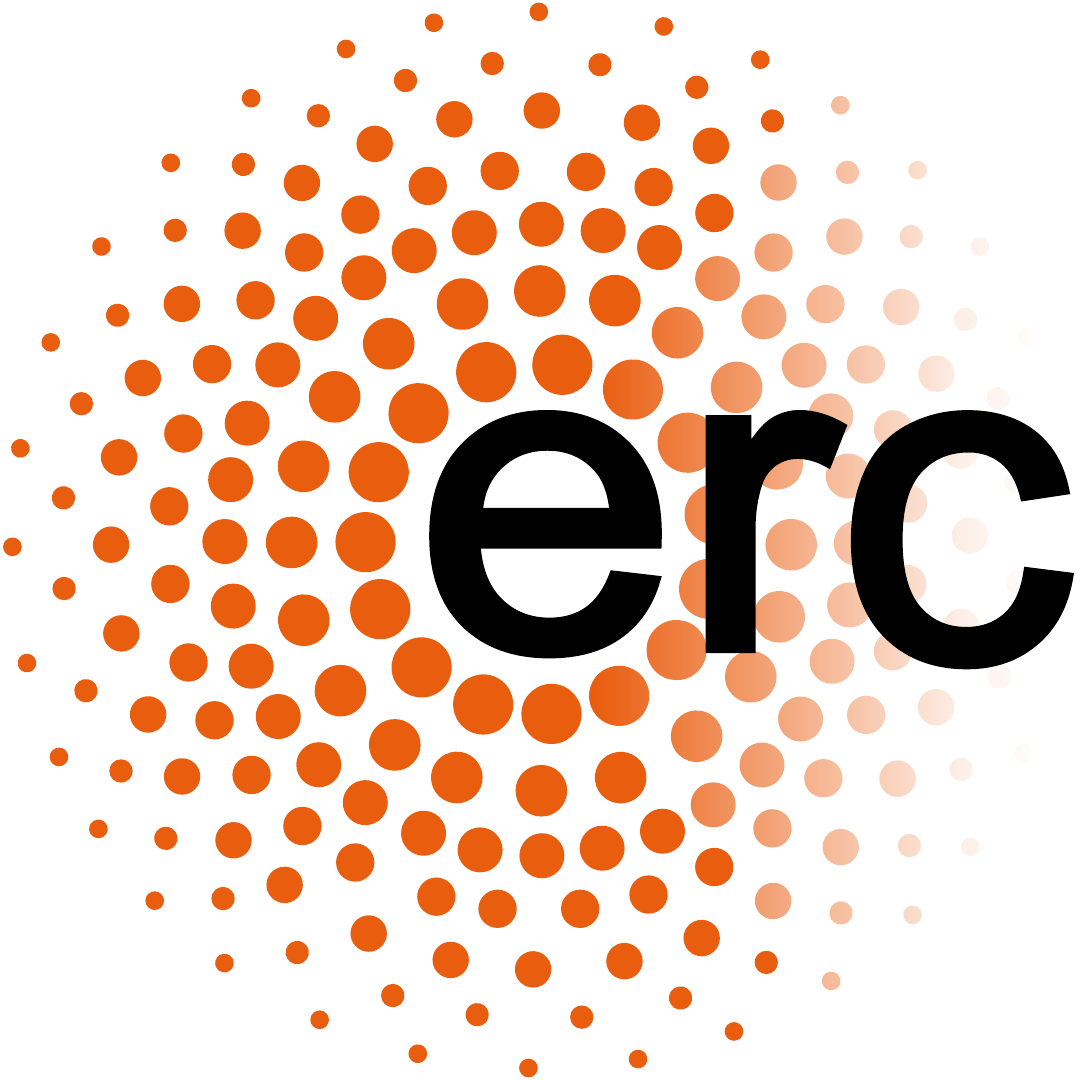} under grant agreement PSAP, number 101002047.
We also wish to acknowledge support from the DEEP-SEA project under grant agreement number 955606.
The Swiss National Science Foundation supports Tal Ben-Nun (Ambizione Project No. 185778). 
The authors would like to thank the Swiss National Supercomputing Centre (CSCS) for access and support of the computational resources.

\bibliographystyle{IEEEtran}
\bibliography{IEEEabrv,references.bib}

\begin{thebibliography}{10}
\providecommand{\url}[1]{#1}
\csname url@samestyle\endcsname
\providecommand{\newblock}{\relax}
\providecommand{\bibinfo}[2]{#2}
\providecommand{\BIBentrySTDinterwordspacing}{\spaceskip=0pt\relax}
\providecommand{\BIBentryALTinterwordstretchfactor}{4}
\providecommand{\BIBentryALTinterwordspacing}{\spaceskip=\fontdimen2\font plus
\BIBentryALTinterwordstretchfactor\fontdimen3\font minus
  \fontdimen4\font\relax}
\providecommand{\BIBforeignlanguage}[2]{{%
\expandafter\ifx\csname l@#1\endcsname\relax
\typeout{** WARNING: IEEEtran.bst: No hyphenation pattern has been}%
\typeout{** loaded for the language `#1'. Using the pattern for}%
\typeout{** the default language instead.}%
\else
\language=\csname l@#1\endcsname
\fi
#2}}
\providecommand{\BIBdecl}{\relax}
\BIBdecl

\bibitem{plasma}
W.~Tang, B.~Wang, S.~Ethier, G.~Kwasniewski, T.~Hoefler, K.~Z. Ibrahim,
  K.~Madduri, S.~Williams, L.~Oliker, C.~Rosales-Fernandez, and T.~Williams,
  ``Extreme scale plasma turbulence simulations on top supercomputers
  worldwide,'' in \emph{Proceedings of the International Conference for High
  Performance Computing, Networking, Storage and Analysis}, ser. SC '16.\hskip
  1em plus 0.5em minus 0.4em\relax IEEE Press, 2016.

\bibitem{cp2k}
\BIBentryALTinterwordspacing
T.~D. K{\"u}hne, M.~Iannuzzi, M.~Del~Ben, V.~V. Rybkin, P.~Seewald, F.~Stein,
  T.~Laino, R.~Z. Khaliullin, O.~Sch{\"u}tt, F.~Schiffmann, D.~Golze,
  J.~Wilhelm, S.~Chulkov, M.~H. Bani-Hashemian, V.~Weber, U.~Borštnik,
  M.~Taillefumier, A.~S. Jakobovits, A.~Lazzaro, H.~Pabst, T.~Müller,
  R.~Schade, M.~Guidon, S.~Andermatt, N.~Holmberg, G.~K. Schenter, A.~Hehn,
  A.~Bussy, F.~Belleflamme, G.~Tabacchi, A.~Gl{\"o}ß, M.~Lass, I.~Bethune,
  C.~J. Mundy, C.~Plessl, M.~Watkins, J.~VandeVondele, M.~Krack, and J.~Hutter,
  ``Cp2k: An electronic structure and molecular dynamics software package -
  quickstep: Efficient and accurate electronic structure calculations,''
  \emph{The Journal of Chemical Physics}, vol. 152, no.~19, p. 194103, 2020.
  [Online]. Available: \url{https://doi.org/10.1063/5.0007045}
\BIBentrySTDinterwordspacing

\bibitem{fmri}
\BIBentryALTinterwordspacing
R.~M. Hutchison, T.~Womelsdorf, E.~A. Allen, P.~A. Bandettini, V.~D. Calhoun,
  M.~Corbetta, S.~Della~Penna, J.~H. Duyn, G.~H. Glover, J.~Gonzalez-Castillo,
  D.~A. Handwerker, S.~Keilholz, V.~Kiviniemi, D.~A. Leopold, F.~de~Pasquale,
  O.~Sporns, M.~Walter, and C.~Chang, ``Dynamic functional connectivity:
  Promise, issues, and interpretations,'' \emph{NeuroImage}, vol.~80, pp.
  360--378, 2013. [Online]. Available:
  \url{https://app.dimensions.ai/details/publication/pub.1051116731}
\BIBentrySTDinterwordspacing

\bibitem{omen}
M.~Luisier, A.~Schenk, W.~Fichtner, and G.~Klimeck, ``Atomistic simulation of
  nanowires in the s p 3 d 5 s* tight-binding formalism: From boundary
  conditions to strain calculations,'' \emph{Physical Review B}, vol.~74,
  no.~20, p. 205323, 2006.

\bibitem{pytorch}
\BIBentryALTinterwordspacing
A.~Paszke, S.~Gross, F.~Massa, A.~Lerer, J.~Bradbury, G.~Chanan, T.~Killeen,
  Z.~Lin, N.~Gimelshein, L.~Antiga, A.~Desmaison, A.~Kopf, E.~Yang, Z.~DeVito,
  M.~Raison, A.~Tejani, S.~Chilamkurthy, B.~Steiner, L.~Fang, J.~Bai, and
  S.~Chintala, ``Pytorch: An imperative style, high-performance deep learning
  library,'' in \emph{Advances in Neural Information Processing Systems 32},
  H.~Wallach, H.~Larochelle, A.~Beygelzimer, F.~d\textquotesingle
  Alch\'{e}-Buc, E.~Fox, and R.~Garnett, Eds.\hskip 1em plus 0.5em minus
  0.4em\relax Curran Associates, Inc., 2019, pp. 8024--8035. [Online].
  Available:
  \url{http://papers.neurips.cc/paper/9015-pytorch-an-imperative-style-high-performance-deep-learning-library.pdf}
\BIBentrySTDinterwordspacing

\bibitem{tensorflow}
\BIBentryALTinterwordspacing
M.~Abadi, A.~Agarwal, P.~Barham, E.~Brevdo, Z.~Chen, C.~Citro, G.~S. Corrado,
  A.~Davis, J.~Dean, M.~Devin, S.~Ghemawat, I.~Goodfellow, A.~Harp, G.~Irving,
  M.~Isard, Y.~Jia, R.~Jozefowicz, L.~Kaiser, M.~Kudlur, J.~Levenberg,
  D.~Man\'{e}, R.~Monga, S.~Moore, D.~Murray, C.~Olah, M.~Schuster, J.~Shlens,
  B.~Steiner, I.~Sutskever, K.~Talwar, P.~Tucker, V.~Vanhoucke, V.~Vasudevan,
  F.~Vi\'{e}gas, O.~Vinyals, P.~Warden, M.~Wattenberg, M.~Wicke, Y.~Yu, and
  X.~Zheng, ``{TensorFlow}: Large-scale machine learning on heterogeneous
  systems,'' 2015, software available from tensorflow.org. [Online]. Available:
  \url{https://www.tensorflow.org/}
\BIBentrySTDinterwordspacing

\bibitem{gt4py}
J.~Dahm, E.~Davis, T.~Wicky, M.~Cheeseman, O.~Elbert, R.~George, J.~J.
  McGibbon, L.~Groner, E.~Paredes, and O.~Fuhrer, ``Gt4py: Python tool for
  implementing finite-difference computations for weather and climate,'' in
  \emph{101st American Meteorological Society Annual Meeting}.\hskip 1em plus
  0.5em minus 0.4em\relax AMS, 2021.

\bibitem{cosmo1}
M.~Baldauf, A.~Seifert, J.~F\"orstner, D.~Majewski, and M.~Raschendorfer,
  ``Operational convective-scale numerical weather prediction with the {COSMO}
  model: Description and sensitivities.'' \emph{Monthly Weather Review,
  139:3387–3905}, 2011.

\bibitem{cosmo2}
\BIBentryALTinterwordspacing
{COSMO}, ``Consortium for small-scale modeling,'' oct 1998. [Online].
  Available: \url{http://www.cosmo-model.org}
\BIBentrySTDinterwordspacing

\bibitem{blas}
L.~S. Blackford, A.~Petitet, R.~Pozo, K.~Remington, R.~C. Whaley, J.~Demmel,
  J.~Dongarra, I.~Duff, S.~Hammarling, G.~Henry \emph{et~al.}, ``An updated set
  of basic linear algebra subprograms (blas),'' \emph{ACM Transactions on
  Mathematical Software}, vol.~28, no.~2, pp. 135--151, 2002.

\bibitem{lapack}
E.~Anderson, Z.~Bai, C.~Bischof, S.~Blackford, J.~Demmel, J.~Dongarra,
  J.~Du~Croz, A.~Greenbaum, S.~Hammarling, A.~McKenney, and D.~Sorensen,
  \emph{{LAPACK} Users' Guide}, 3rd~ed.\hskip 1em plus 0.5em minus 0.4em\relax
  Philadelphia, PA: Society for Industrial and Applied Mathematics, 1999.

\bibitem{berkeley}
\BIBentryALTinterwordspacing
K.~Asanovic, R.~Bodik, J.~Demmel, T.~Keaveny, K.~Keutzer, J.~Kubiatowicz,
  N.~Morgan, D.~Patterson, K.~Sen, J.~Wawrzynek, D.~Wessel, and K.~Yelick, ``A
  view of the parallel computing landscape,'' \emph{Commun. ACM}, vol.~52,
  no.~10, p. 56–67, Oct. 2009. [Online]. Available:
  \url{https://doi.org/10.1145/1562764.1562783}
\BIBentrySTDinterwordspacing

\bibitem{cosma}
G.~Kwasniewski, M.~Kabić, M.~Besta, J.~VandeVondele, R.~Solcà, and
  T.~Hoefler, ``{Red-Blue Pebbling Revisited: Near Optimal Parallel
  Matrix-Matrix Multiplication},'' in \emph{Proceedings of the International
  Conference for High Performance Computing, Networking, Storage and Analysis
  (SC19)}, 2019.

\bibitem{solomonik}
\BIBentryALTinterwordspacing
E.~Solomonik and J.~Demmel, ``Communication-optimal parallel {2.5D} matrix
  multiplication and {LU} factorization algorithms,'' in \emph{Euro-Par 2011
  Parallel Processing}, ser. Lecture Notes in Computer Science, E.~Jeannot,
  R.~Namyst, and J.~Roman, Eds.\hskip 1em plus 0.5em minus 0.4em\relax Springer
  Berlin Heidelberg, 2011, pp. 90--109. [Online]. Available:
  \url{http://dx.doi.org/10.1007/978-3-642-23397-5\_10}
\BIBentrySTDinterwordspacing

\bibitem{optimalStrassen}
G.~Ballard, J.~Demmel, O.~Holtz, B.~Lipshitz, and O.~Schwartz,
  ``Communication-optimal parallel algorithm for strassen's matrix
  multiplication,'' in \emph{Proceedings of the twenty-fourth annual ACM
  symposium on Parallelism in algorithms and architectures}, 2012, pp.
  193--204.

\bibitem{conflux}
G.~Kwasniewski, M.~Kabic, T.~Ben-Nun, A.~N. Ziogas, J.~E. Saethre, A.~Gaillard,
  T.~Schneider, M.~Besta, A.~Kozhevnikov, J.~VandeVondele, and T.~Hoefler, ``On
  the parallel i/o optimality of linear algebra kernels: Near-optimal matrix
  factorizations,'' in \emph{Proceedings of the International Conference for
  High Performance Computing, Networking, Storage and Analysis}, ser. SC
  '21.\hskip 1em plus 0.5em minus 0.4em\relax Association for Computing
  Machinery, 2021.

\bibitem{choleskyQRnew}
E.~Hutter and E.~Solomonik, ``Communication-avoiding {Cholesky}-{QR2} for
  rectangular matrices,'' in \emph{2019 IEEE International Parallel and
  Distributed Processing Symposium (IPDPS)}.\hskip 1em plus 0.5em minus
  0.4em\relax IEEE, 2019, pp. 89--100.

\bibitem{cppapers}
M.~Baskaran, T.~Henretty, B.~Pradelle, M.~H. Langston, D.~Bruns-Smith,
  J.~Ezick, and R.~Lethin, ``Memory-efficient parallel tensor decompositions,''
  in \emph{2017 IEEE High Performance Extreme Computing Conference
  (HPEC)}.\hskip 1em plus 0.5em minus 0.4em\relax IEEE, 2017, pp. 1--7.

\bibitem{tuckerpaper}
V.~T. Chakaravarthy, J.~W. Choi, D.~J. Joseph, X.~Liu, P.~Murali, Y.~Sabharwal,
  and D.~Sreedhar, ``On optimizing distributed tucker decomposition for dense
  tensors,'' in \emph{2017 IEEE International Parallel and Distributed
  Processing Symposium (IPDPS)}.\hskip 1em plus 0.5em minus 0.4em\relax IEEE,
  2017, pp. 1038--1047.

\bibitem{ballard2018communication}
G.~Ballard, N.~Knight, and K.~Rouse, ``Communication lower bounds for
  matricized tensor times khatri-rao product,'' in \emph{2018 IEEE
  International Parallel and Distributed Processing Symposium (IPDPS)}.\hskip
  1em plus 0.5em minus 0.4em\relax IEEE, 2018, pp. 557--567.

\bibitem{gett}
P.~Springer and P.~Bientinesi, ``Design of a high-performance gemm-like
  tensor--tensor multiplication,'' \emph{ACM Transactions on Mathematical
  Software (TOMS)}, vol.~44, no.~3, pp. 1--29, 2018.

\bibitem{tensorGPU}
J.~Kim, A.~Sukumaran-Rajam, V.~Thumma, S.~Krishnamoorthy, A.~Panyala, L.-N.
  Pouchet, A.~Rountev, and P.~Sadayappan, ``A code generator for
  high-performance tensor contractions on gpus,'' in \emph{2019 IEEE/ACM
  International Symposium on Code Generation and Optimization (CGO)}.\hskip 1em
  plus 0.5em minus 0.4em\relax IEEE, 2019, pp. 85--95.

\bibitem{ctf}
E.~Solomonik, D.~Matthews, J.~Hammond, and J.~Demmel, ``Cyclops tensor
  framework: Reducing communication and eliminating load imbalance in massively
  parallel contractions,'' in \emph{2013 IEEE 27th International Symposium on
  Parallel and Distributed Processing}.\hskip 1em plus 0.5em minus 0.4em\relax
  IEEE, 2013, pp. 813--824.

\bibitem{numpy}
\BIBentryALTinterwordspacing
C.~R. Harris, K.~J. Millman, S.~J. van~der Walt, R.~Gommers, P.~Virtanen,
  D.~Cournapeau, E.~Wieser, J.~Taylor, S.~Berg, N.~J. Smith, R.~Kern, M.~Picus,
  S.~Hoyer, M.~H. van Kerkwijk, M.~Brett, A.~Haldane, J.~F. del R{'{\i}}o,
  M.~Wiebe, P.~Peterson, P.~G{'{e}}rard-Marchant, K.~Sheppard, T.~Reddy,
  W.~Weckesser, H.~Abbasi, C.~Gohlke, and T.~E. Oliphant, ``Array programming
  with {NumPy},'' \emph{Nature}, vol. 585, no. 7825, pp. 357--362, Sep. 2020.
  [Online]. Available: \url{https://doi.org/10.1038/s41586-020-2649-2}
\BIBentrySTDinterwordspacing

\bibitem{opt-einsum}
\BIBentryALTinterwordspacing
D.~G. a.~Smith and J.~Gray, ``opt\_einsum - a python package for optimizing
  contraction order for einsum-like expressions,'' \emph{Journal of Open Source
  Software}, vol.~3, no.~26, p. 753, 2018. [Online]. Available:
  \url{https://doi.org/10.21105/joss.00753}
\BIBentrySTDinterwordspacing

\bibitem{dace}
T.~Ben-Nun, J.~de~Fine~Licht, A.~N. Ziogas, T.~Schneider, and T.~Hoefler,
  ``Stateful dataflow multigraphs: A data-centric model for performance
  portability on heterogeneous architectures,'' in \emph{Proceedings of the
  International Conference for High Performance Computing, Networking, Storage
  and Analysis}, 2019, pp. 1--14.

\bibitem{soap}
G.~Kwasniewski, T.~Ben-Nun, L.~Gianinazzi, A.~Calotoiu, T.~Schneider, A.~N.
  Ziogas, M.~Besta, and T.~Hoefler, ``Pebbles, graphs, and a pinch of
  combinatorics: Towards tight i/o lower bounds for statically analyzable
  programs,'' in \emph{Proceedings of the 33rd ACM Symposium on Parallelism in
  Algorithms and Architectures}, 2021, pp. 328--339.

\bibitem{twostep-mttkrp-1}
Q.~Xiao, S.~Zheng, B.~Wu, P.~Xu, X.~Qian, and Y.~Liang, ``Hasco: Towards agile
  hardware and software co-design for tensor computation,'' in \emph{2021
  ACM/IEEE 48th Annual International Symposium on Computer Architecture
  (ISCA)}.\hskip 1em plus 0.5em minus 0.4em\relax IEEE, 2021, pp. 1055--1068.

\bibitem{twostep-mttkrp-2}
K.~Hayashi, G.~Ballard, Y.~Jiang, and M.~J. Tobia, ``Shared-memory
  parallelization of mttkrp for dense tensors,'' in \emph{Proceedings of the
  23rd ACM SIGPLAN Symposium on Principles and Practice of Parallel
  Programming}, 2018, pp. 393--394.

\bibitem{cart-create}
\BIBentryALTinterwordspacing
MPICH, ``Mpi\_cart\_create,'' 2022. [Online]. Available:
  \url{https://www.mpich.org/static/docs/v3.3/www3/MPI\_Cart\_create.html}
\BIBentrySTDinterwordspacing

\bibitem{cart-sub}
\BIBentryALTinterwordspacing
------, ``Mpi\_cart\_sub,'' 2022. [Online]. Available:
  \url{https://www.mpich.org/static/docs/v3.3/www3/MPI\_Cart\_sub.html}
\BIBentrySTDinterwordspacing

\bibitem{P4SC21}
\BIBentryALTinterwordspacing
A.~N. Ziogas, T.~Schneider, T.~Ben-Nun, A.~Calotoiu, T.~De~Matteis,
  J.~de~Fine~Licht, L.~Lavarini, and T.~Hoefler, ``Productivity, portability,
  performance: Data-centric python,'' in \emph{Proceedings of the International
  Conference for High Performance Computing, Networking, Storage and Analysis},
  ser. SC '21.\hskip 1em plus 0.5em minus 0.4em\relax New York, NY, USA:
  Association for Computing Machinery, 2021. [Online]. Available:
  \url{https://doi.org/10.1145/3458817.3476176}
\BIBentrySTDinterwordspacing

\bibitem{tensor-order-nphard}
L.~Chi-Chung, P.~Sadayappan, and R.~Wenger, ``On optimizing a class of
  multi-dimensional loops with reduction for parallel execution,''
  \emph{Parallel Processing Letters}, vol.~7, no.~02, pp. 157--168, 1997.

\bibitem{loopFusionComplexity}
A.~Darte, ``On the complexity of loop fusion,'' in \emph{PACT}, 1999.

\bibitem{ctf-repo}
\BIBentryALTinterwordspacing
{Cyclops Community}, ``Cyclops tensor framework (ctf).'' [Online]. Available:
  \url{https://github.com/cyclops-community/ctf}
\BIBentrySTDinterwordspacing

\bibitem{hptt}
\BIBentryALTinterwordspacing
P.~Springer, T.~Su, and P.~Bientinesi, ``{HPTT}: {A} {H}igh-{P}erformance
  {T}ensor {T}ransposition {C}++ {L}ibrary,'' in \emph{Proceedings of the 4th
  ACM SIGPLAN International Workshop on Libraries, Languages, and Compilers for
  Array Programming}, ser. ARRAY 2017.\hskip 1em plus 0.5em minus 0.4em\relax
  New York, NY, USA: ACM, 2017, pp. 56--62. [Online]. Available:
  \url{http://doi.acm.org/10.1145/3091966.3091968}
\BIBentrySTDinterwordspacing

\bibitem{hptt-repo}
\BIBentryALTinterwordspacing
P.~Springer, ``High-performance tensor transpose library.'' [Online].
  Available: \url{https://github.com/springer13/hptt}
\BIBentrySTDinterwordspacing

\bibitem{hptt-edgar-repo}
\BIBentryALTinterwordspacing
E.~Solomonik, ``High-performance tensor transpose library (forked by edgar
  solomonik).'' [Online]. Available: \url{https://github.com/solomonik/hptt}
\BIBentrySTDinterwordspacing

\bibitem{cutensor}
\BIBentryALTinterwordspacing
Nvidia, ``cutensor,'' 2022. [Online]. Available:
  \url{https://developer.nvidia.com/cutensor}
\BIBentrySTDinterwordspacing

\bibitem{efronbootstrap}
B.~Efron, ``The bootstrap and modern statistics,'' \emph{Journal of the
  American Statistical Association}, vol.~95, no. 452, pp. 1293--1296, 2000.

\bibitem{tiledarray}
\BIBentryALTinterwordspacing
J.~A. Calvin and E.~F. Valeev, ``Tiledarray: A general-purpose scalable
  block-sparse tensor framework.'' [Online]. Available:
  \url{https://github.com/valeevgroup/tiledarray}
\BIBentrySTDinterwordspacing

\bibitem{tblis}
\BIBentryALTinterwordspacing
D.~A. Matthews, ``High-performance tensor contraction without transposition,''
  \emph{SIAM Journal on Scientific Computing}, vol.~40, no.~1, pp. C1--C24,
  2018. [Online]. Available: \url{https://doi.org/10.1137/16M108968X}
\BIBentrySTDinterwordspacing

\bibitem{redblue}
J.~Hong and H.~Kung, ``{I/O} complexity: The red-blue pebble game,'' in
  \emph{STOC}, 1981, pp. 326--333.

\bibitem{externalMem}
J.~S. Vitter, ``External memory algorithms,'' in \emph{European Symposium on
  Algorithms}.\hskip 1em plus 0.5em minus 0.4em\relax Springer, 1998, pp.
  1--25.

\bibitem{redbluewhite}
V.~Elango, F.~Rastello, L.-N. Pouchet, J.~Ramanujam, and P.~Sadayappan, ``Data
  access complexity: The red/blue pebble game revisited,'' Technical Report,
  Tech. Rep., 2013.

\bibitem{redredblue}
J.~E. Savage, ``Extending the hong-kung model to memory hierarchies,'' in
  \emph{International Computing and Combinatorics Conference}.\hskip 1em plus
  0.5em minus 0.4em\relax Springer, 1995, pp. 270--281.

\bibitem{loomisWhitney}
L.~H. Loomis and H.~Whitney, ``An inequality related to the isoperimetric
  inequality,'' \emph{Bull. Amer. Math. Soc.}, vol.~55, no.~10, pp. 961--962,
  10 1949.

\bibitem{irony}
D.~Irony, S.~Toledo, and A.~Tiskin, ``Communication lower bounds for
  distributed-memory matrix multiplication,'' \emph{Journal of Parallel and
  Distributed Computing}, vol.~64, no.~9, pp. 1017--1026, 2004.

\bibitem{hbl1}
T.~M. Smith, B.~Lowery, J.~Langou, and R.~A. van~de Geijn, ``A tight i/o lower
  bound for matrix multiplication,'' \emph{arXiv preprint arXiv:1702.02017},
  2017.

\bibitem{25d}
E.~Solomonik and J.~Demmel, ``Communication-optimal parallel 2.5 d matrix
  multiplication and lu factorization algorithms,'' in \emph{European
  Conference on Parallel Processing}.\hskip 1em plus 0.5em minus 0.4em\relax
  Springer, 2011, pp. 90--109.

\bibitem{petitet}
A.~Petitet and J.~Dongarra, ``Algorithmic redistribution methods for
  block-cyclic decompositions,'' \emph{IEEE Transactions on Parallel and
  Distributed Systems}, vol.~10, no.~12, pp. 1201--1216, 1999.

\bibitem{ibm}
S.~P. Midkiff, ``Local iteration set computation for block-cyclic
  distributions,'' in \emph{Proceedings of the 1995 International Conference on
  Parallel Processing, Urbana-Champain, Illinois, USA, August 14-18, 1995.
  Volume {II:} Software}, C.~D. Polychronopoulos, Ed.\hskip 1em plus 0.5em
  minus 0.4em\relax {CRC} Press, 1995, pp. 77--84.

\bibitem{ecoledesmines}
C.~Ancourt, C.~Fran, and I.~R. Keryell, ``A linear algebra framework for static
  hpf code distribution,'' \emph{A; a}, vol.~1, no.~t2, p.~1, 1993.

\bibitem{rice1}
\BIBentryALTinterwordspacing
K.~Kennedy, N.~Nedeljkovic, and A.~Sethi, ``Efficient address generation for
  block-cyclic distributions,'' in \emph{Proceedings of the 9th International
  Conference on Supercomputing}, ser. ICS ’95.\hskip 1em plus 0.5em minus
  0.4em\relax New York, NY, USA: Association for Computing Machinery, 1995, p.
  180–184. [Online]. Available: \url{https://doi.org/10.1145/224538.224558}
\BIBentrySTDinterwordspacing

\bibitem{rice2}
\BIBentryALTinterwordspacing
------, ``A linear-time algorithm for computing the memory access sequence in
  data-parallel programs,'' in \emph{Proceedings of the Fifth ACM SIGPLAN
  Symposium on Principles and Practice of Parallel Programming}, ser. PPOPP
  ’95.\hskip 1em plus 0.5em minus 0.4em\relax New York, NY, USA: Association
  for Computing Machinery, 1995, p. 102–111. [Online]. Available:
  \url{https://doi.org/10.1145/209936.209948}
\BIBentrySTDinterwordspacing

\bibitem{rice3}
\BIBentryALTinterwordspacing
K.~"Kennedy, N.~Nedeljkovic, and A.~Sethi, \emph{Communication Generation for
  Cyclic(K) Distributions}.\hskip 1em plus 0.5em minus 0.4em\relax Boston, MA:
  Springer US, 1996, pp. 185--197. [Online]. Available:
  \url{https://doi.org/10.1007/978-1-4615-2315-4\_14}
\BIBentrySTDinterwordspacing

\bibitem{riacs}
\BIBentryALTinterwordspacing
S.~Chatterjee, J.~R. Gilbert, F.~J.~E. Long, R.~Schreiber, and S.-H. Teng,
  ``Generating local addresses and communication sets for data-parallel
  programs,'' in \emph{Proceedings of the Fourth ACM SIGPLAN Symposium on
  Principles and Practice of Parallel Programming}, ser. PPOPP ’93.\hskip 1em
  plus 0.5em minus 0.4em\relax New York, NY, USA: Association for Computing
  Machinery, 1993, p. 149–158. [Online]. Available:
  \url{https://doi.org/10.1145/155332.155348}
\BIBentrySTDinterwordspacing

\bibitem{lsu}
A.~Thirumalai and J.~Ramanujam, ``Fast address sequence generation for
  data-parallel programs using integer lattices,'' in \emph{Proceedings of the
  8th International Workshop on Languages and Compilers for Parallel
  Computing}, ser. LCPC ’95.\hskip 1em plus 0.5em minus 0.4em\relax Berlin,
  Heidelberg: Springer-Verlag, 1995, p. 191–208.

\bibitem{cmu}
J.~M. Stichnoth, ``Efficient compilation of array statements for private memory
  multicomputers,'' CARNEGIE-MELLON UNIV PITTSBURGH PA SCHOOL OF COMPUTER
  SCIENCE, USA, Tech. Rep., 1993.

\bibitem{osu}
S.~K.~S. {Gupta}, S.~D. {Kaushik}, S.~{Mufti}, S.~{Sharma}, C.~. {Huang}, and
  P.~{Sadayappan}, ``On compiling array expressions for efficient execution on
  distributed-memory machines,'' in \emph{1993 International Conference on
  Parallel Processing - ICPP'93}, vol.~2, 1993, pp. 301--305.

\end{thebibliography}

\end{document}